\documentclass[graybox]{svmult}

\usepackage{type1cm}        
\usepackage{makeidx}         
\usepackage{graphicx}        
\usepackage{multicol}        
\usepackage[bottom]{footmisc}

\usepackage{newtxtext}       %
\usepackage{newtxmath}       

\usepackage{hyperref}
\hypersetup{
    colorlinks=true,
    linkcolor=blue,
} 
\usepackage{color}

\newcommand{\ron}{\color{red}} 
\newcommand{\bon}{\color{blue}}

\makeindex             

\begin{document}

\title*{Extrasolar enigmas: 
from disintegrating exoplanets to exoasteroids}
\author{Jan Budaj, Petr Kabath and Enric Palle}
\institute{Jan Budaj \at Astronomical Institute, Slovak Academy of Sciences, 05960 Tatranska Lomnica, Slovakia, \email{budaj@ta3.sk}
\and Peter Kabath \at Astronomcal Institute of Czech Academy of Sciences, Fri\v{c}ova 298, 25165, Ond\v{r}ejov, Czech Republic \email{petr.kabath@asu.cas.cz}
\and Enric Palle \at Instituto de Astrof\'{i}sica de Canarias, Calle V\'{i}a L\'{a}ctea, s/n, 38205 San Crist\'{o}bal de La Laguna, Santa Cruz de Tenerife, Spain \email{epalle@iac.es}}
%
%
\maketitle

\abstract{
Thousands of transiting exoplanets have been discovered to date, thanks in great part to the {\em Kepler} space mission. As in all populations, and certainly in the case of exoplanets, one finds unique objects with distinct characteristics. Here we will describe the properties and behaviour of a small group of `disintegrating' exoplanets discovered over the last few years (KIC 12557548b, K2-22b, and others). They evaporate, lose mass unraveling their naked cores, produce spectacular dusty comet-like tails, and feature highly variable asymmetric transits. Apart from these exoplanets, there is observational evidence for even smaller `exo-'objects orbiting other stars: exoasteroids and exocomets. Most probably, such objects are also behind the mystery of Boyajian's star. Ongoing and upcoming space missions such as {\em TESS} and PLATO will hopefully discover more objects of this kind, and a new era of the exploration of small extrasolar systems bodies will be upon us.
}

\section{Introduction}

The exoplanet science discoveries kicked-in after 1992-1995 when the first 
exoplanets were discovered \cite{1992Natur.355..145W} first around a pulsar
and then a hot Jupiter around a solar type star 51 Peg \cite{mayor95}. The exoplanet 51 Peg b was detected from observations of radial velocities (`RV') from the ground with a 1.92-m telescope located at Observatoire de Haute Provence. 

Later hundreds of exoplanets were discovered using the radial velocity method. In 2000, the first {\em transiting} exoplanet HD209458b, again a Jupiter-sized planet in a close-in orbit, was detected \cite{2000ApJ...529L..45C}. New automated ground-based projects to detect transiting exoplanets were started in the first decade of 21st century. The most successful of such projects to date is the WASP survey\footnote{http://www.superwasp.org} which has discovered about 200 transiting planets (April 2019), and there are a number of other successful ground-based exoplanet surveys as well, such as HAT \cite{2018haex.bookE.111B} or KELT \cite{2018haex.bookE.128P}. 

A real breakthrough came with the launch of the CoRoT space mission in 2006. The CoRoT satellite was a french-led ESA mission carrying a 28-cm aperture telescope equipped with 4 CCD detectors dedicated to asteroseismology and exoplanetary transit detections \cite{2006ESASP1306...33B}. The CoRoT mission was terminated in 2013 and it reported 33 exoplanets which are all fully characterized and thus we know both their masses and radii.   

In 2009 a very successful NASA space mission {\it Kepler} was launched carrying a telescope with a mirror of 1.4-m with a large array of CCD detectors \cite{borucki10}. {\em Kepler}, and later its continuation {\em K2} mission, discovered during their lifetimes from 2009 until 2018 about 4000 transiting exoplanets. {\it Kepler/K2} photometric data likely still contain many more new planetary candidates. However, only a few hundreds of the {\em Kepler/K2} planets have been fully characterized, so that we know their masses and radii. This fact is due to the relative faintness of the {\it Kepler/K2} targets and the difficulty of carrying out ground-based follow-up spectroscopic RV observations. However, despite these limitations, {\it Kepler/K2} was able to deliver extremely interesting candidates, among them low mass and rocky planets in the habitable zone such as Kepler-62f \cite{2013Sci...340..587B}, ultra-short period planets such as Kepler-78b \cite{2018NewAR..83...37W}, and multiple planetary systems \cite{2011Natur.470...53L}. Also, new types of objects such as Boyajian's star \cite{boyajian16} and `disintegrating' planets \cite{2018AJ....156..227C} were found with {\it Kepler/K2}. 

In the following text, we will focus on the physics behind the more recently discovered enigmatic objects such as disintegrating and evaporating planets. A significant number of such objects are also expected to be discovered with the most recent and upcoming missions like {\em TESS} and later {\em PLATO}. Before discussing the physics of the disintegrating objects, we briefly introduce the observing strategies which led to the discoveries of these interesting types of exo-objects.

In Section \ref{obs} we describe the methods and observing strategies used to discover or characterize these `dusty objects'. Section \ref{sec:1} contains a crash course on the dust properties which are important to understand the content of this chapter. Sections \ref{exoplanet}, \ref{minor} describe the most interesting disintegrating exoplanets and minor bodies in exoplanetary systems. The special case of Boyajian's star is discussed in the Section \ref{tabby}.
Finally, Section \ref{future} deals with ongoing and future space missions which may bring new fascinating discoveries and open a new era in the study of these extrasolar objects. For a reference, another recent review of disintegrating exoplanets can be found in \cite{lieshout18}. 
  
\section{Observing methods and strategies}
\label{obs}

The most successful methods of exoplanet detection are the transit
and radial velocity measurements. Both methods benefit from their combination, and, in general, all planets detected
by the transit method need follow-up radial velocity measurements for mass
determination. Therefore, all exoplanetary transit space missions try to ensure that the targets in their 
prime sample can be followed-up spectroscopically from the ground.

\subsection{Radial velocities (RV)}

\begin{figure}
\sidecaption[t]
\includegraphics[width=7cm]{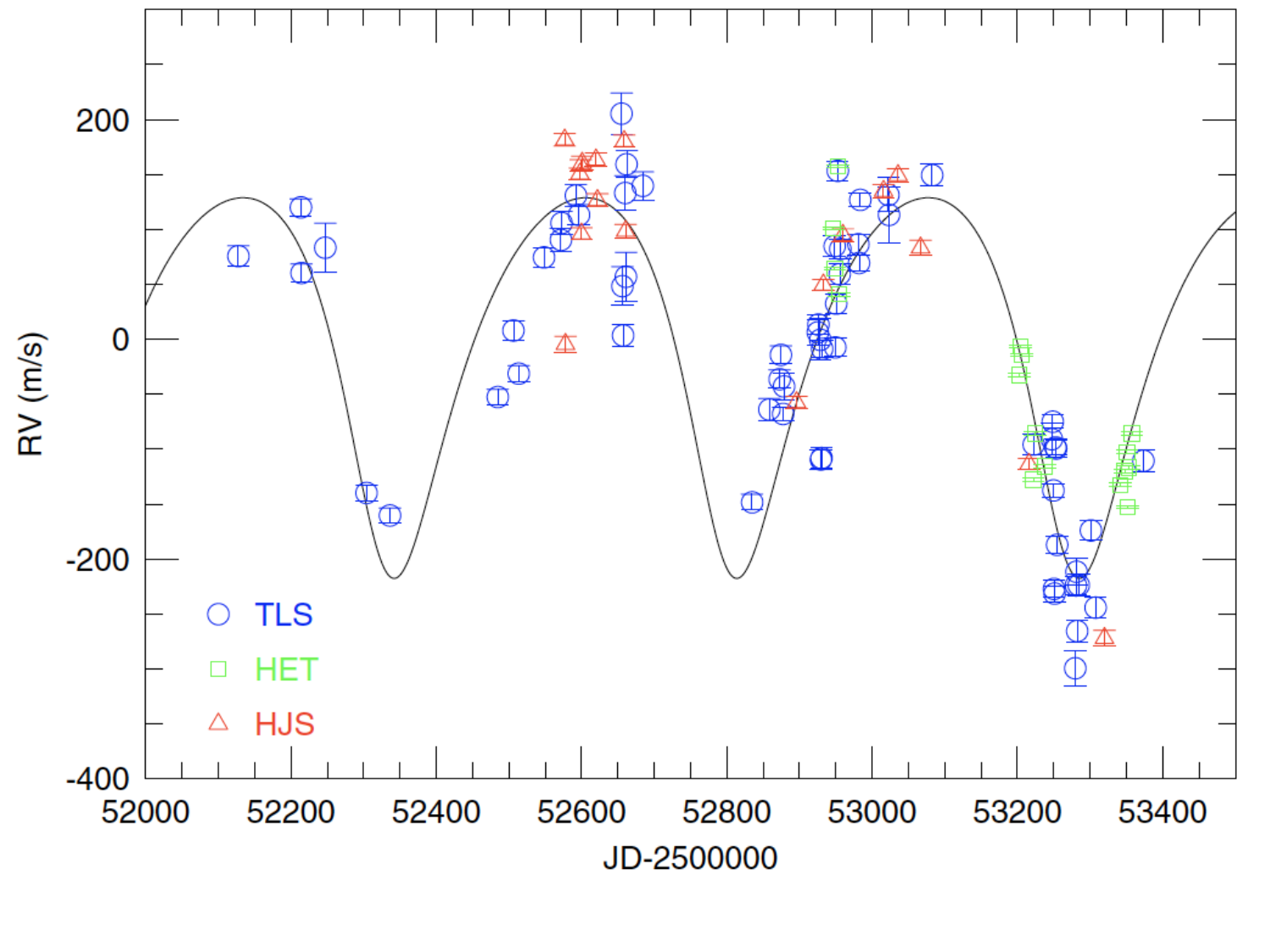}
\caption{Figure shows a typical RV curve of a gas planet obtained with various telescopes around the globe. Figure Credit:\, \cite{2005A&A...437..743H} DOI:\href{	https://doi.org/10.1051/0004-6361:20052850}{10.1051/0004-6361:20052850}, reproduced with permission \textcopyright\,ESO.}\label{rvs}
\end{figure}

The method of discovering and characterizing exoplanets by precise radial velocity measurements is based on Kepler's laws. If the system consists of a star and a planet, these orbit around their common center of mass causing the star to move toward and away from the observer with a given radial velocity that is a function of the mass of the planet. Detailed derivations of the expression for the semi-amplitude $K$ of the radial velocity curve can be found in numerous publications \cite{2010exop.book...27L, 2018haex.bookE...4W}; therefore, we limit ourselves here to only presenting the final expression for the semi-amplitude of the radial velocity curve $K$:
\begin{equation}
K=\frac{1}{\sqrt{1-e^2}} \left(\frac{2\pi G}{P_{\rm orb}}\right)^{1/3} \frac{M_{\rm plan}\sin i}{(M_{\rm star}+M_{\rm plan})^{2/3}} 
\end{equation}
where $G$ is the gravitational constant, $P_{\rm orb}$ the orbital period, $M_{\rm star}$ the stellar mass, $M_{\rm plan}$ the planetary mass, $i$ planetary orbital inclination angle, and $e$ the eccentricity of the planetary orbit. As can be seen from the above equation, the resulting radial velocity and the corresponding semi-amplitude $K$ can be obtained from the observed spectroscopic time series that adequately samples the orbital phases. However, this method can not provide a determination of the inclination, $i$, of the planetary orbital plane. Therefore, the value of planetary mass $M_{\rm plan}$ obtained from the RV measurements is only a lower limit since the value of $i$ is unknown without making use of the photometric transit data. One example of an RV curve is illustrated in Fig.\ref{rvs}

The typical radial velocity semi-amplitude of a large gas planet is of order of tens to hundreds of m/s. On the other hand a typical radial velocity signature of an Earth-sized planet can be as low as few cm/s.


\subsection{The transit method}


If a planet passes in front of the stellar disc along observer's line of sight, then one can observe a periodic dimming of the stellar light, i.e., a transit. Typically a photometric time series with good sampling is obtained a few hours before, then during the transit, and finally a few hours after the transit ends. The basics of the method have been described in great detail elsewhere \cite{winn10}. Here we limit ourselves to expressing the transit depth, $\delta$, as:
\begin{equation}
\delta \propto \frac{\Delta F}{F}=\frac{R_{\rm plan}^2}{R_{\rm star}^2}
\end{equation}
where $\Delta$F is the observed change of flux during a transit, $F$ the flux of the star, $R_{\rm plan}$ the planetary radius, and $R_{\rm star}$ the stellar radius. An advantage of this method is that it can be used to determine the inclination of the planet's orbital plane if the stellar parameters of the host star are known. It is clear that the photometric transit method needs to be combined with spectroscopic observations of a given system in order to fully characterize the exoplanet.
 
\begin{figure}
\sidecaption[t]
\includegraphics[scale=0.5]{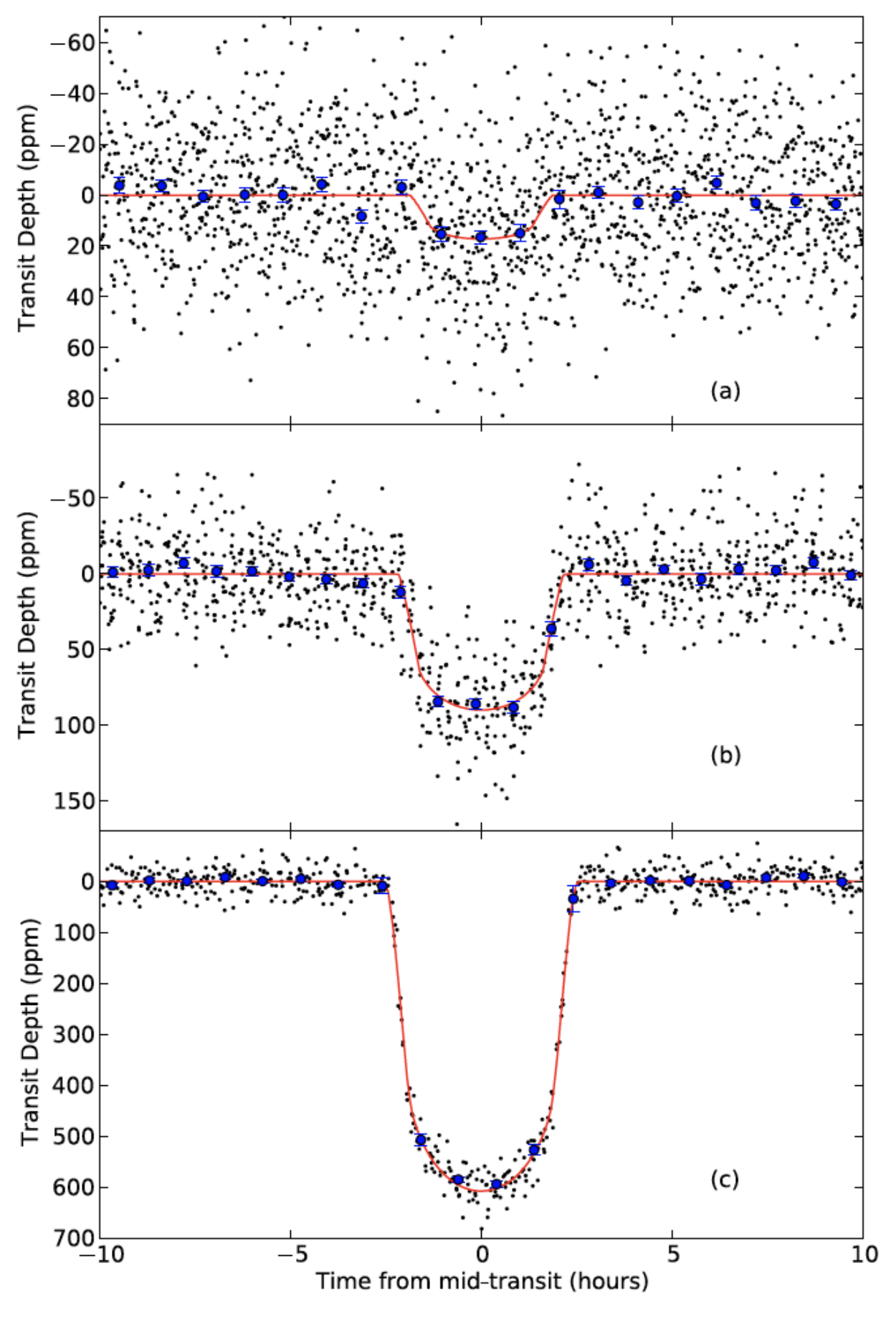}\caption{Figure shows a comparison of light curves obtained with {\it Kepler} for various sized exoplanets from the system Kepler-37, with the smallest being Kepler-37b (upper panel). Reprinted by permission from Springer Nature: Nature, \cite{2013Natur.494..452B}\, DOI: \href{https://www.nature.com/articles/nature11914}{ 10.1038/nature11914}, \textcopyright\,2013. }\label{lcs}
\end{figure}

The detection of hot-Jupiters can be accomplished even with small-aperture telescopes as the typical transit depth, $\delta$, due to a transit of a hot-Jupiter is a few percent of the stellar flux for a main sequence dwarf star. However, the detection of Earth-sized planets requires ultra-precise photometry, typically measured in parts per million (`ppm').
CoRoT-7b was the first example of a small rocky exoplanet showing a transit depth of only a few hundred ppm \cite{2009A&A...506..287L}. The smallest exoplanet currently known to orbit a solar-like star is Kepler-37b \cite{2013Natur.494..452B} and it was discovered by the transit method. Its light curve along with the light curves of two other larger planets in the system are shown in Fig. \ref{lcs}.   \newline

\subsection{Transmission spectroscopy and exo-atmospheres}

Over the past decade, the characterization of exo-atmospheres has started to gain in importance. The first detection of sodium in the exo-atmosphere of a gas giant HD209458b was made from space with the Hubble Space Telescope (HST) \cite{2002ApJ...568..377C}, followed by the spectroscopic detection, also with HST, of an extended hydrogen atmosphere for the same planet \cite{2002ApJ...568..377C,2003Natur.422..143V}. Ground based detection of exo-atmospheres with the transmission spectroscopy method using high spectral resolving power succeeded nearly six years later when sodium was detected in the atmosphere of HD 189733b \cite{2008ApJ...673L..87R}. 


Transmission spectroscopy uses the basic idea that, during a transit, the stellar light has to pass through the exo-planetary atmosphere which forms a thin annulus around the planet. If the atmosphere contains an absorber, such as sodium or any other species, the radius of the planet appears larger at the corresponding wavelength as the species blocks the stellar light. 

When using transmission spectroscopy, typically, a time series of spectra with low spectral resolving power is recorded before, during, and after the transit. Each of the observed spectra from the time series is split into defined photometric bands and then the resulting spectrophotometric light curves are produced and evaluated. The variation in transit depth in the different spectral bands provides information on the absorbing species. This method has successfully confirmed atmospheres for a handful of planets. A metal rich atmosphere was confirmed for the Neptune-sized exoplanet GJ1214b from the ground \cite{2010Natur.468..669B,2011ApJ...743...92B}, followed by many other detections for predominantly gas planets \cite{2019A&A...622A.172M, 2018A&A...616A.145C}. Lately, reports of elements other than sodium and hydrogen have been reported, such as lithium and perhaps a first detection of TiO features \cite{2018A&A...616A.145C,2017Natur.549..238S}. However, this method also has potential application to rocky planets around late type dwarfs \cite{2011ApJ...728...19P} that are recently discovered by TESS and will be found later by PLATO and ELT from the ground.


A slightly different approach is to use a spectrograph with high resolving power. A spectroscopic time series is again obtained on either side of, and during, the transit.  In this case the actual spectra from in- and out-of-transit phase are directly compared. Before a search for planetary atmosphere signatures can start, a careful analysis of the telluric features in the spectra has to be performed and, if necessary, telluric features are removed \cite{2013A&A...557A..56A}. Furthermore the Rossiter-McLaughlin effect which can affect the planetary signal needs to be taken into account \cite{2016A&A...588A.127C}. Regions of prominent lines, such as the sodium doublet (NaD), or potassium region as well as hydrogen lines are typically investigated. The ratio of in- and out-of-transit spectra can reveal a planetary signature \cite{2008ApJ...673L..87R, 2008A&A...487..357S, 2015A&A...577A..62W, 2019A&A...623A.166S, 2019AJ....158..120Z, 2017A&A...608A.135C} as the in-transit spectra also possess an excess signal from the planetary atmosphere.


\subsection{Observing strategies of exoplanetary space missions}

The detection of exoplanets is most efficient from space with the transit method. Therefore, we will introduce the observing strategies and principles of such missions. Different space missions dedicated to the search of planets via transit detection have followed different observing strategies. The first one, CoRoT, monitored several fields for a series of long (150 days) and short (30 day) periods. On the contrary, the space mission {\em Kepler} monitored a single field for 4 years. The selected field in the region of the Cygnus and Lyra constellations contained more than 150,000 stars \cite{borucki10} that were monitored. This part of the {\it Kepler} mission yielded about 2000 exoplanets and several thousand candidates. In 2013 the {\it Kepler} team needed to adopt a different observing strategy due to problems with the spacecraft gyroscopes. The mission was renamed {\em K2} and it observed one field for typically 70 days and then pointed towards a new field.  Over the ensuing four years, the {\em K2} mission yielded about 1000 exoplanets and several hundred additional candidates \cite{2014PASP..126..398H}. There were numerous interesting discoveries among these missions, and many  ``firsts" reported, such as: the circumbinary planet Kepler-16b \cite{2019ApJ...875L...7D}, the oldest known multiplanet system Kepler-444 \cite{2015ApJ...799..170C}, the first {\em Kepler} rocky planet Kepler-10b \cite{2011ApJ...729...27B}, and the first planet with a radius smaller than the Earth \cite{2012Natur.482..195F}. The {\em K2} mission was retired in late 2018 when the fuel was depleted. 

However, {\it Kepler} also discovered a new class of `disintegrating'
planets. In the following text, we lay the theoretical ground for understanding these highly enigmatic planets among the known types of exoplanetary systems.

\section{Dust environment in exoplanets}
\label{sec:1}

In this section we introduce the basic physical properties of astrophysical dust which
will be important for understanding the subsequent sections.
At sufficiently low temperatures and high density, grains of 
condensates can be formed out of a gas phase. 
Such grains are usually called ``dust'', although some authors use 
the more generic term ``condensates''. 
At the same time, the term ``grain'' often includes not only solid 
grains but also liquid droplets.
Such condensates are usually confined to ``clouds''.
These can not only be clouds in the atmospheres of cool objects 
but also vast interstellar dust clouds.

\begin{figure}
\sidecaption[t]
\includegraphics[width=7.5cm,angle=0]{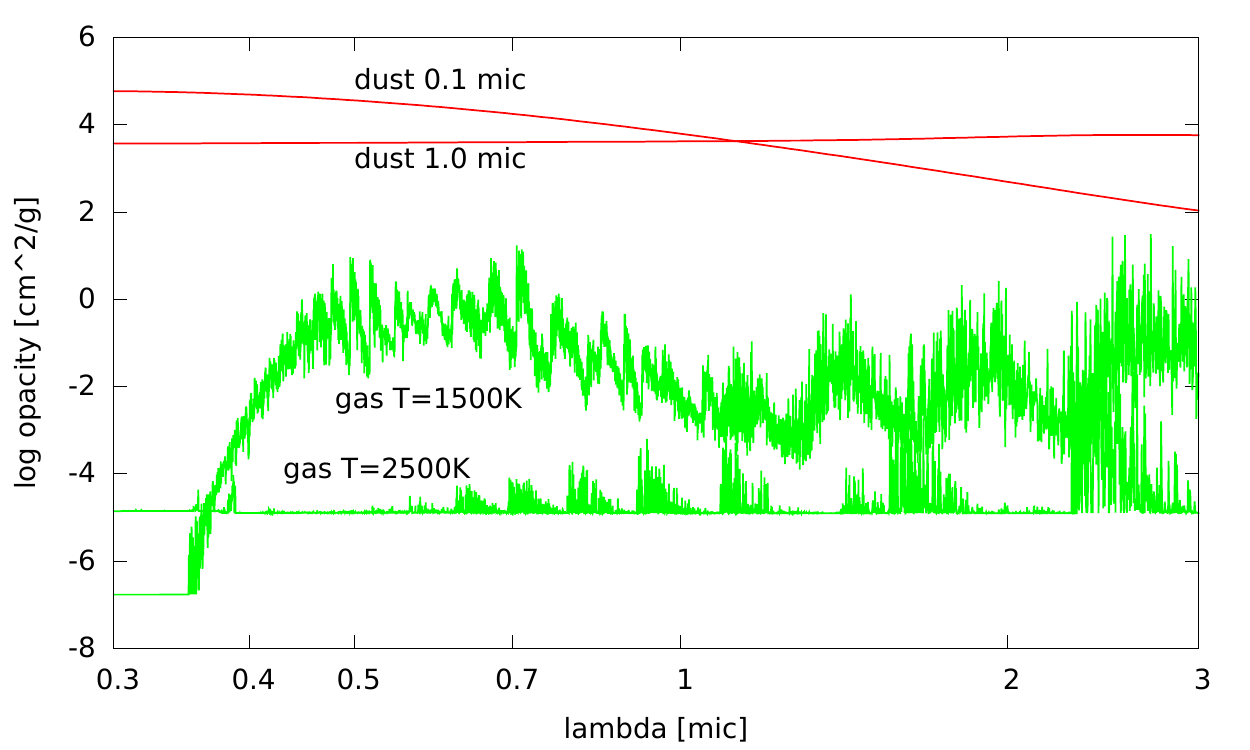}
\caption{Comparison between the gas opacities at two temperatures
and dust opacity of forsterite for two particle sizes.}
\label{opac}
\end{figure}

The reason why dust is so important for our objects will become 
obvious from the following everyday experience.  Our atmosphere contains
water. If this water is in the form of a gas one can easily see distant mountains which are 100 km away. However, once the water condenses and forms clouds or a fog, the visibility can drop to 10 meters or even less. 
Thus the opacity, which is a measure of the non-transparency of the material (see Sec.\ref{xsection}), could be much higher if the material were in the form of dust rather than gas.
Figure \ref{opac} illustrates the opacity of gas and dust in the visible 
and near infrared regions per gram of material.
The opacity of the gas in this example is based on an assumed solar chemical composition and 
a density $\rho=10^{-14}$ g\,cm$^{-3}$ \cite{marigo09}. 
For the dust opacity we used the illustrative mineral forsterite with a particle size 
of about 0.1 and 1 micron \cite{budaj15}.
It should be mentioned that, as a rule, not all the gas can turn into
a condensate. For solar composition material, dust can account for 
roughly 1\% of the mass.
Still, as can be seen from the figure, the dust opacity will easily 
overtake that of the gas.

\subsection{Absorption, Scattering, and Extinction}

The optical properties of condensates may not only influence, but
fully govern, the emerging spectrum and even the structure of a dusty object. 
Dust can absorb the impinging radiation and convert it directly into
heating the grains. This process is called `absorption'
or `true absorption' to emphasize that the photon is destroyed or thermalized.
It is quantified by the absorption opacity. 

Dust can also scatter the radiation in a process called `scattering'. Scattering  mainly changes the direction
of the photon without significantly affecting its energy. So the scattered radiation
is somewhat decoupled from the medium and flows through and around it without heating it. 
This process is characterized by the scattering opacity.
Furthermore, scattering can be highly anisotropic, a property that is  
described by means of the phase function, which depends on 
the scattering angle (the deflection angle from the original direction
of the impinging radiation).
The most prominent feature is a strong forward scattering peak for large
values of the so-called `scaled particle size', $X=2\pi a/\lambda$ where $a$ is the particle size (radius)   
and $\lambda$ is the wavelength of the radiation.

The combined effect of absorption and scattering is referred to as the `extinction'.
Finally, formation of dust can also affect the chemical composition of an object. It removes the condensed elements from the gas phase
within the dust cloud. Subsequently, various processes and forces may
decouple gas and dust, creating chemical inhomogeneities. 

Absorption and scattering by large particles (relative to the wavelength, i.e., large $X$) is wavelength independent. However, {\em scattering} by small particles has a very strong, $\lambda^{-4}$, dependence (Rayleigh scattering) and {\em absorption} by small particles has a $\lambda^{-1}$ dependence. Blue light is scattered and attenuated more efficiently, and for this reason dust generally causes a reddening of the light passing through a dust cloud. The extinction at some wavelength (or filter)
in magnitudes is the difference between the observed and intrinsic brightness: $A(V)=V_{\rm obs}-V_{\rm int}$. Reddening (selective extinction/color excess) is usually expressed as a difference between the observed and intrinsic color index:
\begin {equation}
E(B-V)=(B-V)_{\rm obs}-(B-V)_{\rm int}=A(B)-A(V)
\end {equation}
A relative slope of the wavelength dependence of the extinction can be characterized by a single parameter $1/R(V)$ where R(V) is \cite{cardelli89}:
\begin {equation}
R(V)=\frac{A(V)}{E(B-V)}=\frac{A(V)}{A(B)-A(V)}
\end {equation}
$R(V)$ is sensitive to the particle size. The typical value of $R(V)$ for interstellar dust in our Galaxy is $3.1\pm0.2$. 
The absolute amount of the extinction as a function of $\lambda$ (the extinction curve) can be characterized by two parameters: $R(V)$ and $E(B-V)$.

\subsection{Cross-section, Opacity, and Phase Function}
\label{xsection}

The optical properties of the dust are given by the complex index of refraction
of the material it is made from (which is a function of wavelength), and 
further depends on the size and shape of the particles. 
These properties of the grains are often expressed in the form of 
cross-sections for absorption, scattering, and extinction 
$C_{\rm a}, C_{\rm s}, C_{\rm e}$, respectively.
The cross-sections are related to the projected area of the dust particles of radius $r$ via efficiency factors $Q_{\rm a}$, $Q_{\rm s}$, and $Q_{\rm e}$, 
for absorption, scattering, and extinction, respectively:
\begin {equation}
C_{\rm a}=Q_{\rm a}\pi r^{2},~~~C_{\rm s}=Q_{\rm s}\pi r^{2},~~~
Q_{\rm e}=Q_{\rm a}+Q_{\rm s}.
\end {equation}
The cross-sections are related to the absorption and scattering opacities,
$\kappa_{\nu,a}$ and $\kappa_{\nu,s}$, of the condensates at radiation frequency, $\nu$, by:
\begin {equation}
\kappa_{\nu,\{a,s\}} \equiv C_{\nu,\{a,s\}}/m_g = \frac{3}{4} \frac{Q_{\nu,\{a,s\}}}{\rho_g r}
\label{ks}
\end {equation}
where $\kappa$ is the cross section per unit mass and has the dimensions of cm$^2$ g$^{-1}$, $m_g$ is the mass of a dust grain, $\rho_g$ is the bulk material density of the grain, and we have assumed spherical particles for simplicity.  Note that $\kappa$ is nearly exclusively a property of the material and may not depend at all on the mass in dust grains per unit volume of the medium, $\bar{\rho}$.  Finally, for completeness, we note that the quantity $\alpha \equiv \bar{\rho} \kappa$ is defined as the linear extinction coefficient (with units cm$^{-1}$), and may be useful in certain circumstances.

%
%
%
Using these opacities, the monochromatic optical depth due to scattering and absorption along the line of sight $z$ is then given by:
\begin {equation}
\tau_{\nu}=  \int \bar{\rho}(z)
\left[\kappa_{\nu,a}(z) + \kappa_{\nu,s}(z)\right] dz.
\end {equation}
The sum of the absorption and scattering opacities is referred to as a total opacity.  For the special idealized case of single-size dust particles, this reduces to
\begin {equation}
\tau_{\nu}= \frac{3}{4} \frac{\left(Q_{\nu,a}+Q_{\nu,s}\right)}{\rho_g r} \int \bar{\rho}(z) dz.
\end {equation}
We can see from this, that for a fixed amount of dust mass per unit volume of the medium, i.e., $\bar{\rho}$ = constant, the optical depth would become monotonically larger with decreasing particle size, as $1/r$.  However, in Mie scattering, once the particle size becomes substantially less than the radiation wavelength, $\lambda = c/\nu$, then the $Q$ factors for the cross section drop dramatically, and the optical depth stops rising with further decreases in particle size.  This is the reason why, for observing wavelengths in the visible, it is often stated that particle sizes comparable to a micron are the most efficient at blocking light.  

\begin{figure}
\sidecaption[t]
\includegraphics[angle=0,width=7.5cm]{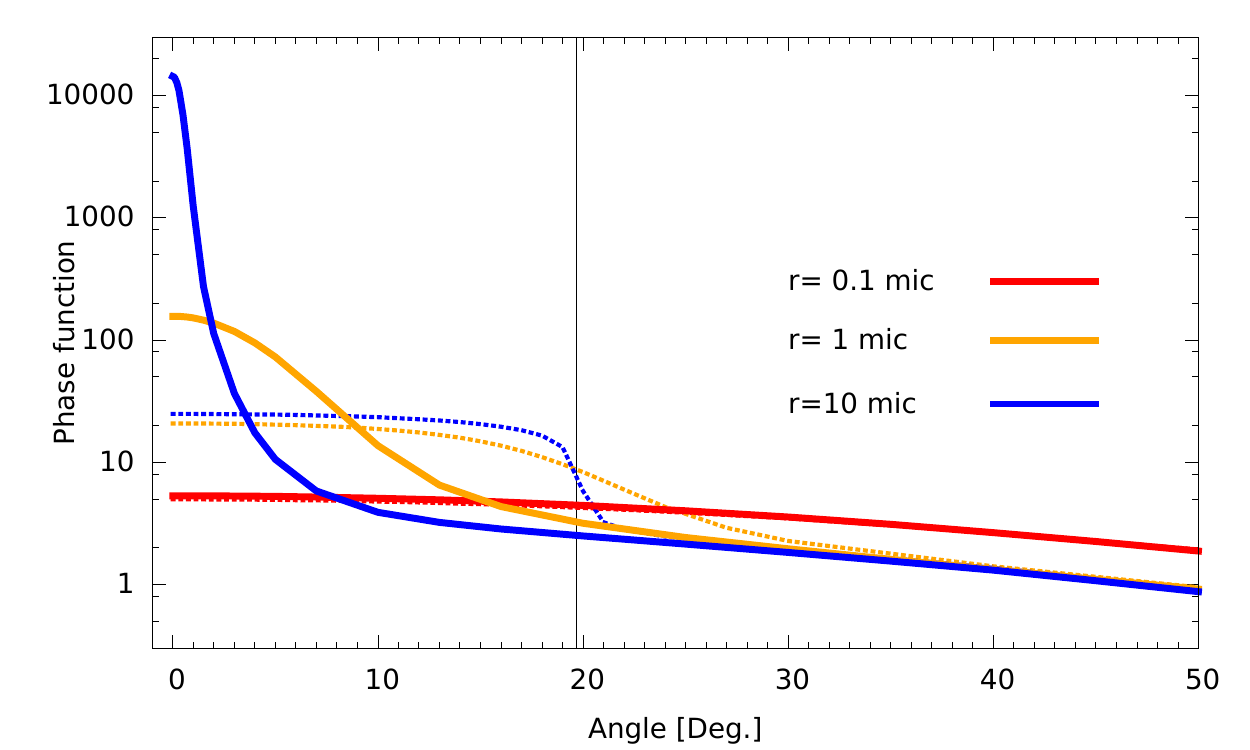}
\caption{
Phase functions assuming a point source of light  
(solid) versus the phase functions assuming
the finite dimension of the stellar disc (dotted).
Example is for enstatite at 600 nm for different dust particle radii.
The vertical line illustrates the angular radius of the stellar
disc of WASP-103 as seen from the planet WASP-103b.
}
\label{pfbf} 
\end{figure}

The angular distribution of the scattered light
is described by the phase function,
$p(\theta)$, where $\theta$ is the scattering angle which
measures the deflection of the scattered photon from its original direction.
The phase function is normalized such that its integral over all  
solid angles is $4\pi$.
An example of the dust phase function at small angles is displayed in 
Figure \ref{pfbf}. 
One can see a strong increase towards zero phase angle which is called 
the forward scattering peak. The amplitude and width of this peak are quite 
sensitive to the particle size and wavelength. Calculations of phase functions usually assume
an incident parallel beam of light. However, if the dust cloud were very close 
to the star, its angular dimension (as seen by the dust grain) could be comparable to, or wider than, the width of the forward scattering peak. 
This will be the case in our objects and one has to take that into account
\cite{budaj13,devore16,garcia18}.
The same figure also illustrates this effect on dust particles
located in the atmosphere of the exoplanet WASP-103b \cite{gillon14,southworth15}.

It is sometimes useful to define a mean cosine of the scattering angle 
$g$, also known as the asymmetry parameter. It has values from -1 to 1 and is calculated from the phase function:
\begin {equation}
g=\int p(\theta) \cos(\theta) d\Omega ~/4\pi.
\label{g}
\end {equation}

\subsection{Albedo, Equilibrium Temperature, and Radiative Acceleration}

Let's assume that a dust particle is irradiated by its host star with effective temperature $T_{*}$, solid angle $\Omega_*$, and intensity approximated by the Planck function $B_{\nu}(T_*)$.
The particle can scatter some of the light from the star, and we define a quantity called single-scattering albedo, $\varpi$, which describes the reflecting properties of the grains. It is a fraction of the energy which is scattered by the particle:
\begin {equation}
\varpi_{\nu}=
\frac{C_{\nu,s}}{C_{\nu,a}+C_{\nu,s}} ~.
\end {equation}
This scattered light does not heat the particle.
Apart from scattering, the particle can also absorb the stellar radiation at a rate:
\begin{equation}
\Omega_* \int  C_{\nu,a} B_{\nu} (T_*) d\nu
\end{equation}
This energy heats the particle to a temperature $T_g$. 
Subsequently, the grain emits thermal radiation and cools at a rate:
\begin{equation}
4\pi \int  C_{\nu,a} B_{\nu} (T_g) d\nu , 
\end{equation}
A balance between the absorbed and re-radiated 
energy sets the grain equilibrium temperature (provided that the grain is not also sublimating; but see Eqn.~\ref{equil2}).
It can be obtained by solving the radiative equilibrium equation for $T_g$:  
\begin{equation}
\Omega_* \int  C_{\nu,a} B_{\nu} (T_*) d\nu = 4\pi \int  C_{\nu,a} B_{\nu} (T_g) d\nu
\label{equil}
\end{equation}

Assuming that the opacities are grey (i.e., they do not depend of the frequency)
the grain temperature is simple given by:
\begin {equation}
T^{\rm grey}_g=T_{*} \left(\frac{\Omega_*}{4\pi}\right)^{1/4}.
\label{tgrey}
\end {equation}

A dust grain irradiated by a star with effective temperature
$T_{*}$, mass $M_{*}$, radius $R_{*}$, and surface flux $F_{\nu}$ 
experiences a radiative acceleration $a_{\rm R}$.
It is usually expressed as a parameter $\beta$ relative to 
the gravitational acceleration $a_{\rm G}$:
\begin {equation}
\beta=\frac{a_{\rm R}}{a_{\rm G}} = \frac{R_{*}^{2}}{G M_{*} c }
\int \left[\kappa_{\nu,a} + 
(1 - g) \kappa_{\nu,s}\right]
F_{\nu}(T_{*})d\nu.
\end {equation}
where $G$ is gravitational constant, $c$ is speed of light, 
and $g$ is the previously mentioned asymmetry parameter.
Thus, in the two extreme cases of forward vs back scattering
of the stellar radiation, the scattering adds either nothing to the radiative
acceleration or has a factor of 2 enhancement relative to the absorption 
term.

Extensive online tables of such dust properties devoted mainly to 
exoplanets are publicly available \cite{budaj15}.
They are based on codes that calculate cross-sections of dust particles
using Mie theory such as \cite{bohren83} from the complex indices of refraction for specific materials. 
For example, the Heidelberg - Jena - St.Petersburg - Database of
Optical Constants is a very convenient source of refractive index
measurements \cite{henning99,jager03}.

\subsection{Dust Condensation}

Depending on the state quantities, such as temperature and pressure,
matter composed of a single component usually exists in one particular 
phase, e.g. gas, liquid, or solid. 
The Clausius-Clapeyron equation, which gives a relation between the temperature
and pressure, marks a transition or boundaries between the different
phases. 
Once the temperature drops below the condensation temperature
(at a certain pressure) or the pressure exceeds the equilibrium
(saturated) vapour pressure (at a particular temperature)
the dust starts to condense out of the gas.
Condensates may be in either the liquid or solid phase.
The equilibrium vapor pressure where the transition occurs
can be approximated by 
\cite{kimura02,lieshout14}:
\begin {equation}
P_{v}(T)=\exp(-A/T+B)
\end {equation}
where $A,B$ are material-specific sublimation parameters.

Materials with low vapour pressure or high condensation temperature
(refractory materials) condense first out of a hot cooling gas 
(or last to evaporate if the dust were heated). 
For a solar chemical composition
these are mainly calcium and aluminum oxides
such as corrundum (Al$_{2}$O$_{3}$), grossite (CaAl$_{4}$O$_{7}$)
and hibonite (CaAl$_{12}$O$_{19}$).
They are followed by titanium compounds such as perovskite 
(CaTiO$_{3}$) or TiO$_{2}$ at lower temperatures.
The most important refractory species are usually silicates.
They form two branches:
pyroxenes (Mg$_{x}$Fe$_{1-x}$SiO$_{3}$) and olivines
(Mg$_{2y}$Fe$_{2-2y}$SiO$_{4}$). In each branch a fraction of 
magnesium atoms can be replaced by iron.
Iron free pyroxene is called enstatite (MgSiO$_{3}$) while 
an iron free olivine is forsterite (Mg$_{2}$SiO$_{4}$). 
The other extreme member of the olivine family is fayalite 
(Fe$_{2}$SiO$_{4}$). 
Silicates are a type of glass and, as such, are quite transparent 
in the optical region, although they can scatter light quite efficiently. 
The amount of iron can affect
their absorption properties significantly \cite{dorschner95}.
Other refractory dust species which might be encountered in such an
environment are amorphous carbon, graphite (C), 
silicon carbide (SiC), Quartz (SiO$_{2}$),
spinel (MgAl$_{2}$O$_{4}$), or akermanite (Ca$_{2}$MgSi$_{2}$O$_{7}$).
At the other end of the condensation temperature scale
are volatile species such as water and ammonia.
In between are numerous compounds, depending on the chemical 
composition and pressure, for example
sulfides and alkali halides, and troilite but we are not likely to 
observe these in such hot and close disintegrating objects.

Apart from the temperature, the occurrence of 
a particular dust component also depends critically on the abundances 
and the availability of the chemical elements which form 
the compound. The element with the lowest abundance
is typically the limiting factor for the abundance of the whole 
compound.
The solar abundances 
\footnote{Note that the abundances are defined, using the element 
number density $N$, as the number of atoms of an element per $10^{12}$ atoms of hydrogen ($\log N/H+12$). The present-day solar photospheric abundances are generally in a good agreement with the abundances derived from the CI carbonaceous chondrite meteorites,
except for a few elements such as H, He, and Li.} 
of Ca, Al, and Ti are relatively small 6.34, 6.45, and 4.95,
respectively \cite{asplund09}. 
That is why silicates and/or iron dust which are composed of silicon, 
magnesium, and iron with abundances of 7.51, 7.60, and 7.50, 
respectively are usually more abundant and dominate extinction
processes.

\begin{figure}
\sidecaption[t]
\includegraphics[angle=0,width=7.5cm]{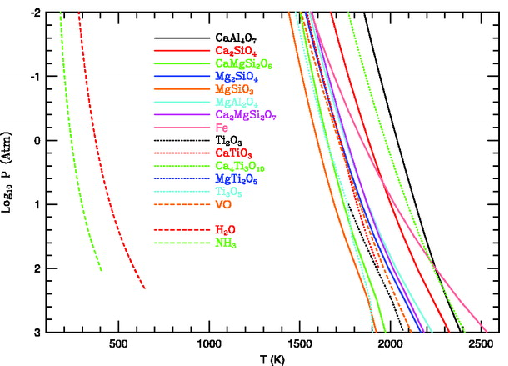}
\caption{
Condensation temperatures of several compounds as a function of 
atmospheric pressure assuming a solar composition gas.
Taken from \cite{burrows06}. Courtesy of ApJ. }
\label{cond}
\end{figure}

The condensation properties of various compounds are nicely summarized 
in Figure \ref{cond}. 
Here the condensation curves are plotted as a function of 
atmospheric pressure. They were calculated mainly for 
the atmospheres of brown dwarfs or giant exoplanets and assume
a solar chemical composition \cite{burrows06} but contain many 
dust species which are also relevant for our objects.

\subsection{Dust Sublimation}

Dust particles are also subject to sublimation 
\cite{kimura02,lieshout14}. 
The mass-loss flux (rate per unit area) from a solid surfaces at temperature $T$ 
in vacuum is 
\begin {equation}
J(T) = \alpha P_{v}(T) \sqrt{\frac{\mu u}{2\pi k_{B}T}},
\end {equation}
where $\alpha$ is the evaporation coefficient,
$P_{v}(T)$ the equilibrium vapor pressure, 
$\mu$ the molecular weight, $u$ the atomic mass unit,
and $k_{B}$  Boltzmann's constant.
The mass-loss rate from a spherical dust grain of mass 
$m_{g}=4\pi r^{3}\rho_{g}/3$ and surface area $S=4\pi r^{2}$ is then
\begin {equation}
\frac{d m_{g}}{d t} = -SJ.
\end {equation}
Taking into account that
\begin {equation}
\frac{d m_{g}}{d r} = S\rho_{g}
\end {equation}
the change in the particle radius is given by:
\begin {equation}
\frac{d r}{d t} =\frac{d r}{d m_{g}}\frac{d m_{g}}{d t} = 
-\frac{J}{\rho_{g}}.
\end {equation}
Sublimation represents a phase transition which consumes
heat and cools the particle. 
If that heat is not negligible one has to take it into account
in computing the equilibrium temperature of the grain.
In such a case the energy absorbed by the particle per unit time is balanced by the energy radiated by the particle plus the heat consumed for the phase transition. Equation \ref{equil} then reads
\begin{equation}
\Omega_* \int  C_{\nu, a} B_{\nu} (T_*) d\nu
= 4\pi \int C_{\nu, a} B_{\nu}(T) d\nu -\mathcal{L}\frac{d m_{g}}{d t}
\label{equil2} 
\end{equation}
where $\mathcal{L}$ is the latent heat of sublimation per unit mass.
The characteristic timescale for sublimation is
\begin{equation}
\tau =\frac{m_{g}}{|d m_{g}/d t|}.
\end{equation}

\section{Known Disintegrating Exoplanets}
\label{exoplanet}

The great majority of exoplanets that we know of were discovered by the transit method.
Nominal planet transits are symmetric and periodic without any significant variations in their 
shape or depth over time. This changed in 2012 when a strange object named Kepler-1520-b was discovered \cite{rappaport12}.

\subsection{Kepler-1520b}

\begin{figure}
\sidecaption[t]
\includegraphics[angle=0,width=6.cm]{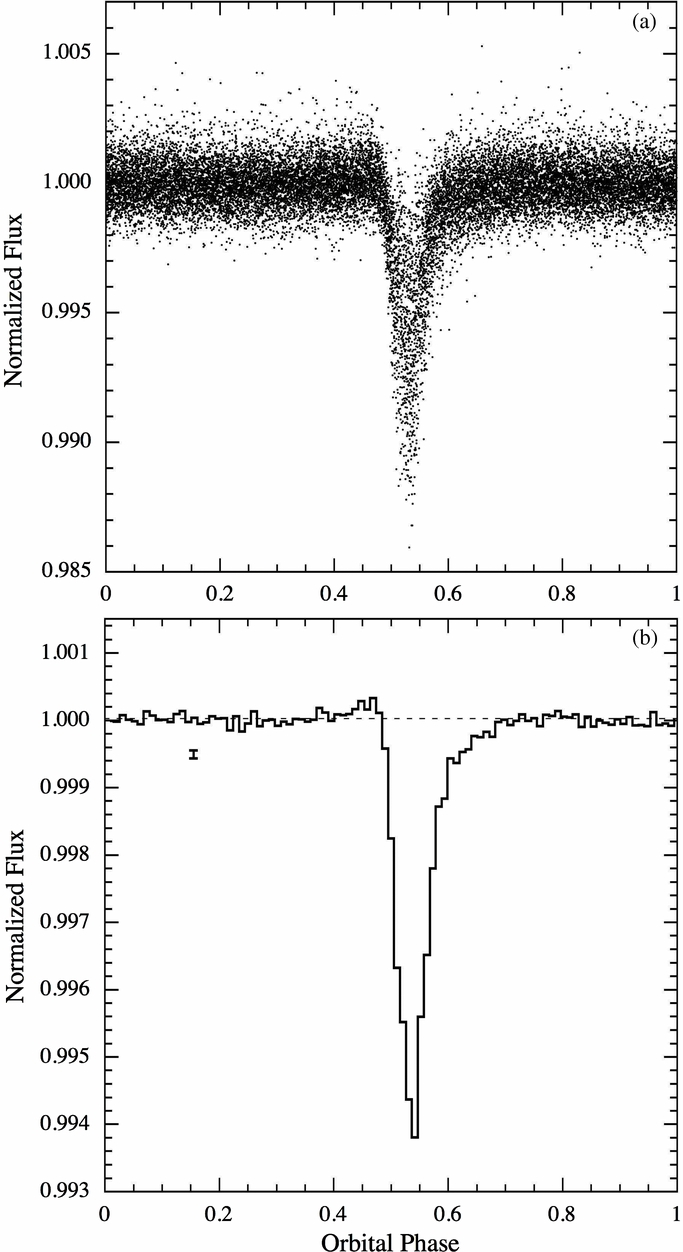}
\caption{
Top: First long cadence Kepler observations of KIC 1255 folded with the 15.7-hr orbital period.
Bottom: Binned and averaged light curve.
Taken from \cite{rappaport12}. Courtesy of ApJ.}
\label{kic1255a}
\end{figure}

Kepler-1520b is an exoplanet also known as KIC 12557548b (KIC1255b).
It became a prototype of a very rare new class of exoplanets called Disintegrating Exoplanets. It was found in the {\em Kepler} data.
The host star is a V=16 mag main sequence K4V type star. Its effective temperature, mass, and radius are about $T_{\rm eff}=4440$\,K, $M=0.7M_{\odot}$, and $R=0.65 R_{\odot}$, respectively
\cite{rappaport12,schlawin18}.
The star is active and has spots which cause $\sim$1\% variability with a period of about 22.9 days which enabled its rotation period to be determined \cite{budaj13,kawahara13}.
In its light curve, the discoverers noticed something like transits but
they were highly variable, sometimes as deep as 1.2\%, sometimes even missing.
The strictly periodic transit signal had a very short period of about 15.7 hours.
Figure \ref{kic1255a} illustrates the observed data folded with this period
which yields the average light curve. One can see a significantly increased spread of fluxes in the points 
during the transit indicating the variability in the transit depth.
Another interesting feature becomes obvious from the binned and averaged light curve (bottom panel of Fig.~\ref{kic1255a}).
It is highly asymmetric and features a steeper ingress and slower egress.
The strict periodicity and short period of the transits indicate that they may be caused by some body orbiting the star on a very close orbit.
The fact that the transits are sometimes missing implies that the body itself is very small,
smaller than the Earth, otherwise it would be detected in every transit.
Follow-up radial velocity measurements did not detect any reflex motion of the star
which puts an upper limit on the mass of the body of 89 $M_{\oplus}$
\cite{croll14,masuda18} which places the body deep into the planetary regime.
However, what is then causing the variable asymmetric transits?

\subsubsection{Interpretation}
The interpretation it that a body on such tight orbit around the star is heated
to about 2000 Kelvin. At such temperatures even rock melts and can evaporate which may drive a thermal wind off the surface \cite{rappaport12,perez13}.
Gas escapes the planet at a rate larger than 0.1 $M_{\oplus}$/Gyr dragging dust grains with it. Alternatively, the dust may condense 
out of the gas when it cools during or after escape from the planet. The mixture of gas and dust expands beyond the Hill sphere radius of the planet. It flows ``down hill" out of the potential well of the planet through the L1 point towards the star or via the L2 point away from the star. Strong radiative forces on the dust cause a weakening of the effective gravity which drives the dust into higher orbits that lag progressively behind the planet. It is this dust which is causing the transits
and this is also the reason why we observe a steep ingress followed by a gradual egress.

Once such a fine dust cloud forms around and behind the planet it may not be
stable and is prone to variability.
For example when the dust cloud is thin the planet surface is intensively
irradiated, which leads to more evaporation, outflows, and condensation, 
thereby producing more dust. In turn, the thick dust cloud shields the planet 
and the evaporation drops, limiting the production of dust, and the cloud 
dissipates. This limit cycle can apparently operate, even on a timescale from 
orbit to orbit, but there are longer intervals of order a week where the transits
are reduced to a level where they are not detected (see also the following section 
on the variability).

Producing and maintaining a substantial outflow of gas and dust is relatively 
simple in bodies with the surface gravities of asteroids, where the thermal speed
of the material exceeds the escape speed (see, e.g., Fig.\,8 of \cite{lieshout18}).
For at least some common minerals the vapour pressure at $\sim$2000 K is sufficiently high that for bodies below lunar size, the direct Jeans' escape mass loss rates could exceed that required to produce the inferred dust rates in KIC 1255b of $\sim$1 $M_\oplus$/Gyr.  For more substantial bodies, e.g., Mercury, 
Mars, and Earth, a Jeans' outflow of 1 $M_\oplus$ per Gyr of heavy molecules 
becomes nearly impossible \cite{rappaport12,perez13,lieshout18}.
For such massive bodies, a different escape mechanism has been proposed, namely
a Parker-type hydrodynamic wind \cite{rappaport12,perez13}. Roughly speaking this requires thermal speeds 
that are only $\sim$1/4 of the escape speed in order to work \cite{rappaport12,perez13}.
One issue with the requirement of a planet losing 1 $M_\oplus$ per Gyr is that 
if it has of only $\sim$ $10^{-3}-10^{-2} \, M_{\oplus}$, then it will have a lifetime of only 1-10 Myr. If the lifetime of the host star is measured in Gyr, then the a prior probability of seeing one of its planets in that evaporative state are rather low. However, obviously if one surveys a large number of stars, then the odds of seeing a few such systems is non-negligible. 
Since such planets may have lost most of their mass their observations
open a unique window into planetary interiors and their chemical composition \cite{bodman18,ridden18}.

\subsubsection{Variability}
It was mentioned above that the transits are variable.
They vary on a very short timescale from one orbit to another, i.e., in less than one day.
This variability is strong, sometimes more than a factor of 2 from one orbit to the next one, and appears to be stochastic and associated with the deep core of the transit \cite{rappaport12,werkhoven14}. 
However, a modulation of the transit depth was also found that appears to be  
anti-correlated with the periodic rotational variability (22.9 days) of the stellar flux\cite{kawahara13,croll15}.

There is also a smooth long-term variability in the egress part of the light curve associated with the dust tail on timescales of about 1.3 yr which is not seen in the core of the transit \cite{budaj13,schlawin18}.
There might also have been a period of decreased activity, i.e., when the transits were shallower on average, during  2013-2014 \cite{schlawin16}.
This longer-term variability in the depth and shape of the transits indicates that the dust cloud associated with the planet may not be homogeneous and has at least two components; an inner tail (or coma) and an outer tail which may behave
differently (e.g., when subjected to magnetic fields or stellar winds) or have different properties (particle size, chemical composition)
\cite{budaj13,werkhoven14}. On the contrary, \cite{lieshout16} arrived at the conclusion that, as far as the pure shape of the average transit profile is concerned, it is well reproduced in their calculations and there is no need to invoke two such constituents.
A similar long-term variability of the transit, namely a monotonic decrease of its depth over the four-year duration of the {\em Kepler} mission was found in another disintegrating exoplanet, KOI 2700b \cite{rappaport14}.

The reason for the above mentioned long-term variability has not been well studied but it has been argued that it may be associated with the magnetic activity of the star and be analogous to the comet tail disconnection events observed in some of the comets in our Solar System \cite{budaj13,kawahara13}.
However, it has also been argued that the modulation of the transit depth with stellar rotation may be due to occultations of the stellar spots rather than the magnetic activity \cite{croll15}.

\subsubsection{Pre-transit brightening}
There is a very interesting tiny feature in the transit light curve, barely visible
in Figure \ref{kic1255a}. It is a small brightening just before the transit,
already noted by the discoverers \cite{rappaport12},
which is referred to as a pre-transit brightening.
It is not due to the star getting brighter. It is caused by the scattering properties
of the dust. As shown in Figure \ref{pfbf}, the dust does not scatter the light
isotropically but mainly in the forward direction. For the same reason a driver gets 
blinded when the Sun is near, but not in, the driver's immediate field of view, but the windshield is dirty and this nonetheless scatters the sunlight into his/her eyes.

\begin{figure}
\sidecaption[t]
\includegraphics[angle=0,width=7.5cm]{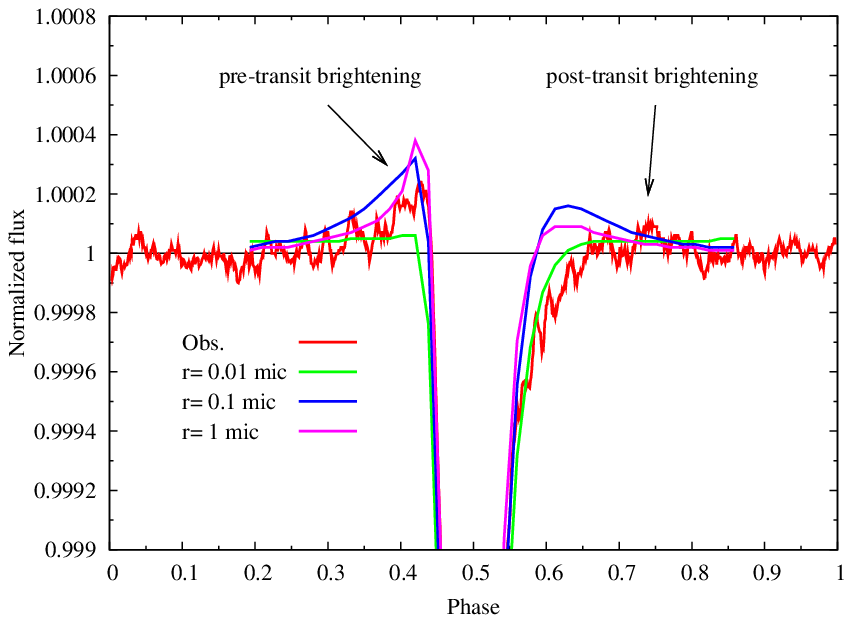}
\caption{
{\em Kepler} light curve of KIC 1255 (red) zoomed so that the pre-transit brightening is clearly visible. Models (green, blue, purple) demonstrate that this feature is sensitive to the particle size. Taken from \cite{budaj13} and reproduced with permission 
\textcopyright ESO.}
\label{kic1255b}
\end{figure}

In our system, this happens mainly in the vicinity
of the transit. While we cannot identify this light during or after the transit, 
since it is overlaid with the ongoing absorption, we can see it just before the transit
(Figure \ref{kic1255b}). This feature is sensitive to the particle size
and it enables us to estimate that the size of particles in the tail is about 0.1-1 micron.
At the same time this effect confirms that the transit events are caused by a dusty 
tail passing in front of, and close to, the star.
Apart from these features in direct transits, the forward scattering effect can, in principle, be used to detect non-transiting dusty-tailed exoplanets by searching for positive bumps in the light curves \cite{devore16,garcia18}. 

\subsubsection{Particle size and chemical composition}
A number of authors have studied the {\em Kepler} light curves of KIC 1255b attempting to derive 
the chemical composition and grain size distribution of the transiting dust material
 \cite{brogi12,budaj13,werkhoven14}.
This problem is partially degenerate and one can fit such monochromatic\footnote{In this context `monochromatic' means transits that are observed in only a single waveband.} transits
with different chemical composition and particle size.
The pre-transit brightening is sensitive to the particle size and the observed brightening indicates
particles 0.1-1 micron in size.
On the other hand, the length of the tail is highly sensitive to the sublimation properties of the grains. Corundum and 0.2-5 micron grains are most favoured for this reason and the mass loss rate amounts to 0.6-16 Earth masses per Gyr \cite{lieshout14,lieshout16}.

\begin{figure}
\sidecaption[t]
\includegraphics[angle=0,width=7.5cm]{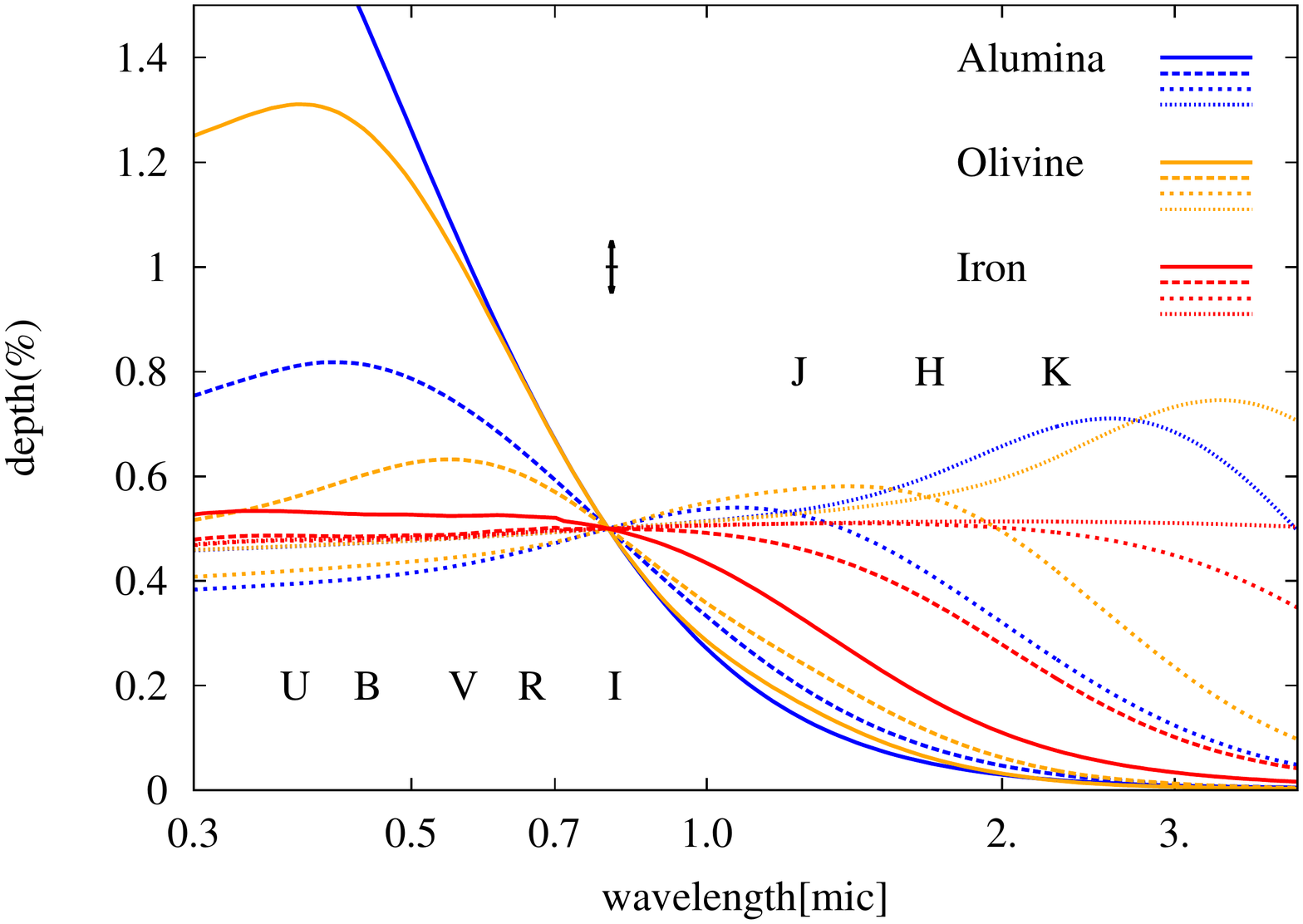}
\caption{
Theoretical transit depths for three species alumina, olivine, and iron; each for 
the particle size of 0.1 (solid), 0.16 (dashed), 0.4 (short-dashed), 1.0 (dotted) micron
as a function of wavelength. Depth is normalized such that the transit in the I filter is about 0.5\% deep.
}
\label{depth}
\end{figure}
More information and a deeper insight can be achieved with multi-wavelength observations.
This is because the opacity of dust changes with the wavelength and the behaviour is different for grains of different chemical composition and size.
Consequently, under the assumption that the tail is optically thin, the transit depth would depend
on the wavelength, the particle size, and the chemical composition.
This is illustrated in Figure \ref{depth} which shows theoretical transit depths for three species: corundum (alumina), olivine, and iron. 
One can see that the transits produced by small particles of corundum or
silicates would be much deeper at the shorter wavelengths. This is because
scattering dominates extinction and scattering on small particles
(relative to the wavelength) is approaching the Rayleigh regime with a strong $\lambda^{-4}$ dependence. Extinction by large particles is almost grey. 
The problem is that the observations must be carried out at different wavelengths simultaneously because of the above mentioned strong variability of the transit depths. 

Such observations in the optical and near-infrared regions have not detected a significant difference in the transit depth across these wavelengths. This implies that the dust particle size must be larger than $\sim$0.5 micron.  In a scenario where dust grains are lifted directly from the surface of the planet, this in turn implies that the planet should be less massive than Mercury otherwise its gravity would prevent such direct dust ejection \cite{croll14,schlawin16}.  
However, as mentioned above, dust might have also condensed later beyond the potential well of the planet.
Additional multi-wavelength observations in $z^{'}, g^{'}, u^{'}$ filters indicate slightly
larger depths at shorter wavelengths and particle sizes of about 0.25-1 micron \cite{bochinski15}.
Recent 3D models of the dust dynamics including the sublimation and 3D radiative transfer pointed out the possibility that the tail may be optically thick. In this case
the transit depth might be constant with the wavelength even for smaller particles
and mass loss rates may reach 80 Earth masses per Gyr \cite{ridden18}.

Apart from KIC 1255b, two other systems of this kind have been discovered,
KOI 2700b \cite{rappaport14} and K2-22b \cite{sanchis15}.
The first one is similar in its transit profile to KIC1255b, and the latter system is described in more detail below.

\subsection{K2-22b}

This exoplanet (also known as EPIC201637175B) was discovered with the {\em Kepler} follow-on mission ({\em K2}) by \cite{sanchis15}.
It is in some respects similar to KIC 1255b. The host star is cooler and smaller. It is an M0V type red dwarf (r = 15.01 mag) with effective temperature, mass, and radius of about $T_{\rm eff}=3830$\,K, $M=0.6\,M_{\odot}$, and $R=0.57 \, R_{\odot}$, respectively. The host star rotates with a period of 15.3 days and has a `close' (3 magnitudes fainter) companion, separated by about $2^{\prime\prime}$.
The planet K2-22b is smaller than 2.5 $R_{\oplus}$, is less massive than 1.4 $M_{J}$, and has a very short orbital period of only $9.145872 \pm 0.000024$ hours. It is losing mass in the form of a dusty tail at a rate $\approx 2 \times 10^{11} g\, s^{-1}$ \cite{sanchis15}. 

\begin{figure}
\sidecaption[t]
\includegraphics[angle=0,width=5.0cm]{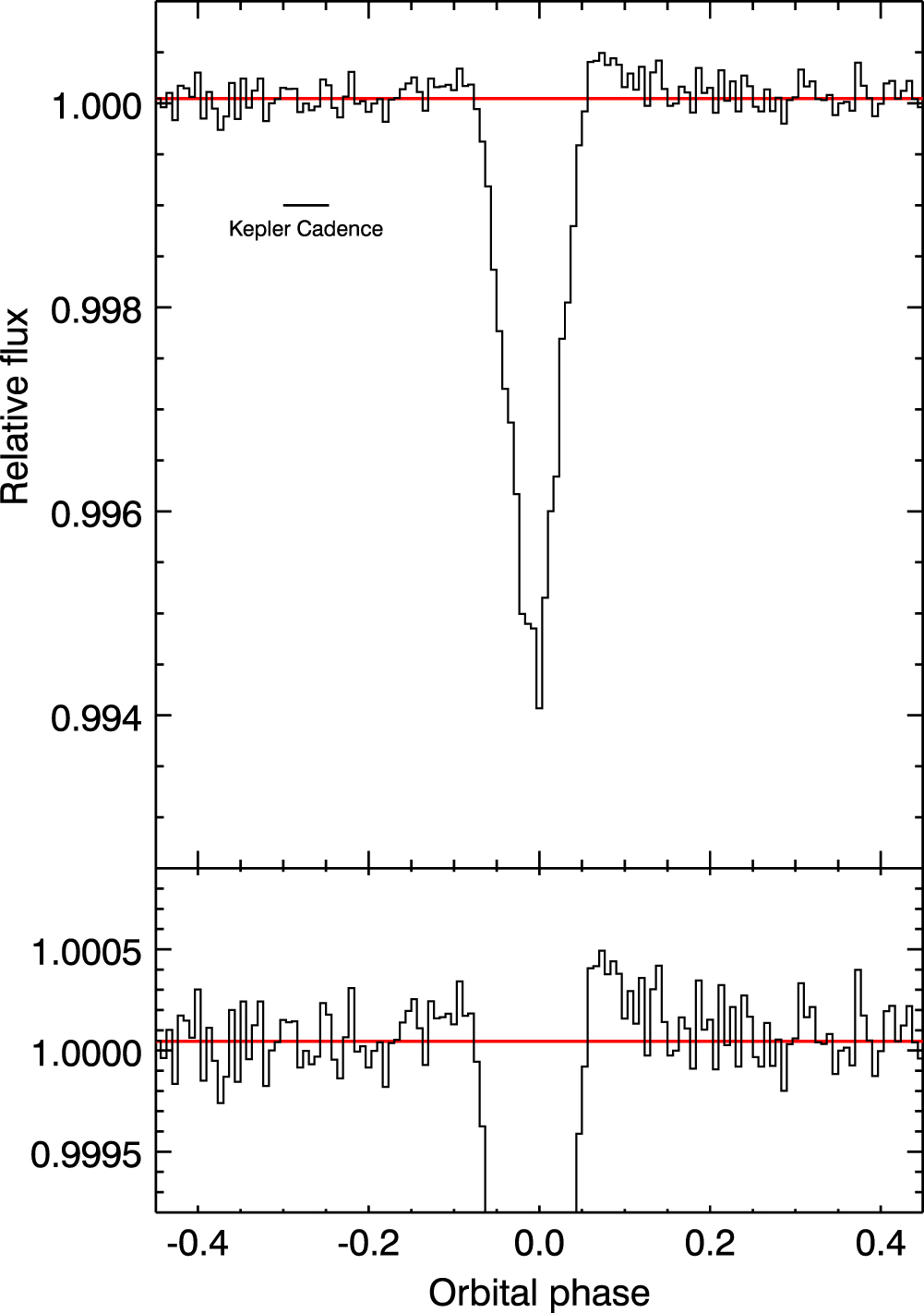}
\caption{
Binned and averaged {\em K2} light curve of K2-22 folded with the orbital period.
It features a post-transit brightening likely indicative of a leading dust tail.
Taken from \cite{sanchis15}. Courtesy of ApJ.}
\label{k2-22}
\end{figure} 
As in the case of KIC 1255b, the transits are asymmetric and highly variable. They are on average about 0.5\% deep but the depth changes from 0 to 1.3\% from transit to transit. The duration of the transits is about 50 minutes. 
The average transit shape is shown in Figure \ref{k2-22}. The special feature of this exoplanet is that it exhibits a post-transit brightening. 
Based on the lesson learned from KIC 1255b, this likely indicates that the planet also has a dusty tail but it is pointing in the opposite direction. In other words, the planet is orbiting the star with its dust tail heading forward.
This is most probably due to the host star being colder and fainter than KIC 1255. Its radiation does not exert sufficient pressure on the dust grains to force them into a higher orbit, and thereby trail the star.
Thus, the dust can flow from the planet toward the L1 point and the host star, and then descend into the potential well of the star. Since the Keplerian velocity of these orbits is higher, 
these grains overtake the planet and form a leading dust tail \cite{sanchis15}.

The follow up multi-wavelength transit observations with the GTC in the visible region found no evidence for a wavelength dependence in three out of the four transits observed \cite{sanchis15,colon18}. One transit, however, did indicate that 
the transit depth is greater at the bluer wavelengths. This sets an upper limit on the dust grains of about 0.4-0.6 micron. The forward scattering peak indicates particle sizes of about 0.5 micron.
Although the dust is the major opacity source, the gas might be detected 
in the cores of some strong spectral lines such as NaI in high resolution spectra. \cite{gaidos19} searched for such gas absorption during the transits
of K2-22b and Kepler-1520b but did not detect any spectral signatures.

\section{Minor bodies in Extrasolar systems}
\label{minor}

As in our solar system, minor bodies are also expected to exist in extrasolar systems. While we do not yet have the capability to detect structures similar to the main asteroid belt or the Oort cloud in other planetary systems, the first extrasolar minor bodies have recently been detected.

\subsection{Exo-Asteroids: a debris disk around WD 1145+017}
\label{exoasteroid}

In the solar system, asteroids are defined as minor bodies in the inner solar system that show significant departures from spherical shape dictated by hydrostatic equilibrium. The first extrasolar minor bodies were discovered by {\em K2} \cite{2015Natur.526..546V} in the form of disintegrating material orbiting the white dwarf WD 1145+017. 

It has long been known that some white dwarfs have dusty debris disks around them
\cite{2010ApJ...722..725Z,2014A&A...566A..34K}, and also that many of them
(about 1/4 - 1/2) have heavy elements in their atmospheres that should have already sunk into the stellar interiors, unless they were replenished by infalling orbiting material \cite{2003ApJ...584L..91J,2007ApJ...663.1285J, 2006ApJ...646..474K}.  Vanderburg et al \cite{2015Natur.526..546V} observed for the first time this process in action by detecting a white dwarf being transited by `at least one and likely multiple disintegrating planetesimals with periods ranging from 4.5 hours to 4.9 hours'. The detected transits are marked by being asymmetric, and even irregular, with respect to normal transiting planets, indicating that they do not correspond to solid spherical bodies, and can be as deep at 55 per cent.  In addition, most of the observed transits are much longer in duration than the $\sim$1-2 min expected transit time of an asteroid with an orbital period of 4.5 hour.
Note that the white dwarf radius and luminosity are quite low.
Its apparent brightness, luminosity, distance, effective temperature, mass, and radius are about $g=17.0$ mag, $L=0.0093 L_{\odot}$, $d=174$ pc, $T_{\rm eff}=15900$\,K, $M=0.6\,M_{\odot}$,  and $R=1.4\, R_{\oplus}$, respectively. The orbital period of 4.5h corresponds to 
a distance of about $1 R_{\odot}$ from the star.
A combination of this distance and stellar luminosity yields equilibrium temperatures 
of about 1400-1700 K which is similar to those of disintegrating planets. 

\begin{figure}
\sidecaption[t]
\includegraphics[angle=0,width=5.0cm]{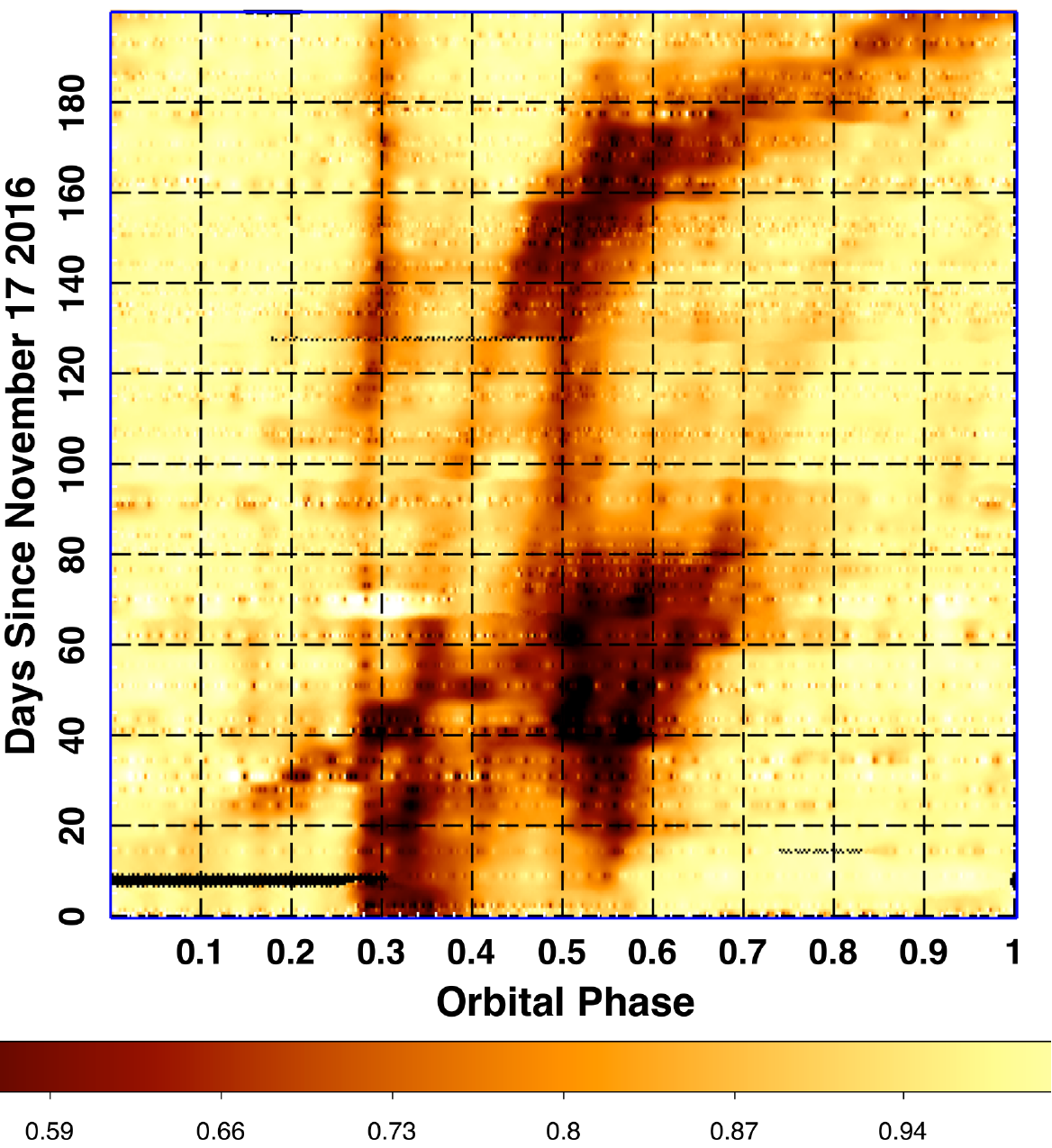}
\caption{
Waterfall diagram of WD1145 phased with the base period of 4.49126 days. 
Objects with the base period follow the vertical line while objects with 
different periods crisscross the diagram on different tracks.
Taken from \cite{2018MNRAS.474..933R} by permission of Oxford University Press.}
\label{wd1145water}
\end{figure}

This object has attracted the attention of exoplanet observers.
Croll et al. \cite{2017ApJ...836...82C} conducted ground and space follow-up observations on WD 1145+017. The observations  confirmed that the white dwarf is orbited by multiple short-period objects, that egress times were longer than ingress times, and the duration of the transits was longer than expected, pointing again to cometary tail-like structures behind the debris fragments.
These asteroids are nicely visualized with a 'waterfall' diagram presented in
Rappaport et al. \cite{2018MNRAS.474..933R} showing the evolution of 
the phase light curve (see Fig.\ref{wd1145water}). One can easily identify
several objects with slightly different periods crisscrossing the picture. 

\begin{figure}
\includegraphics[angle=0,width=10.0cm]{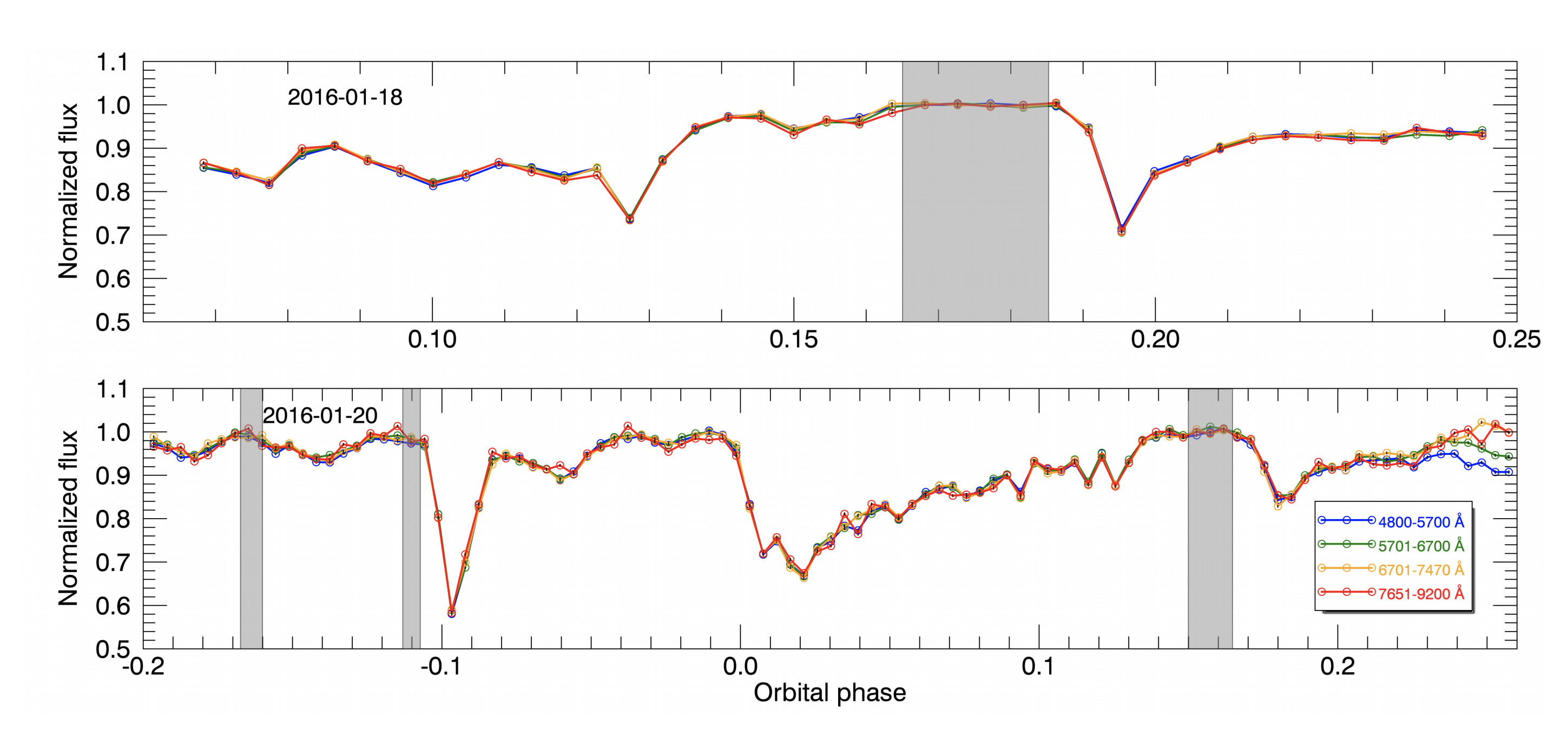}
\caption{
GTC light curves of WD 1145+017 taken simultaneously in four wavebands and covering several dips. The nearly identical dip profiles in the four bands can be used to constrain the dust grain sizes to larger than 0.5 $\mu$m. The divergence of the curves after phase 0.22 in the lower panel is due to atmospheric effects. Adapted from \cite{2016A&A...589L...6A}
 and reproduced with permission \textcopyright ESO.}
\label{wd1145}
\end{figure}

Croll et al. \cite{2017ApJ...836...82C} also did not detect any transit chromaticity.
Alonso et al \cite{2016A&A...589L...6A} and Izquierdo et al \cite{2018MNRAS.481..703I} used the 10-m GTC telescope to check for chromaticity but found the transits to be gray over the optical range from 480 to 920 nm (see Figure \ref{wd1145}), indicating that particle sizes smaller than 0.5 micron can be excluded. 
From their observations, Alonso et al \cite{2016A&A...589L...6A} concluded that the radius of single-size particles in the tail materials must be $\approx$ 0.15 $\mu$m or larger, or $\approx$ 0.06 $\mu$m or smaller. They also report low amplitude variations in the light curves suggesting that dusty material is continuously passing in front of the stellar disk. 

Zhou et al. \cite{2016MNRAS.463.4422Z} and Xu et al \cite{2018MNRAS.474.4795X} also observed these dips in multiple photometric bands in the visible and infrared. They find no difference in the transit depths once infrared observations are corrected for excess emission from a dusty disk. Xu et al \cite{2018MNRAS.474.4795X} conclude that there must be a deficit of small particles in the transiting material and that only large particles can survive without sublimating at the effective temperatures prevalent at these short orbital periods.

Xu et al. \cite{2019AJ....157..255X} found the first detection of chromaticity, showing that UV transit depths are always shallower than those in the optical. They proposed a model to explain this observations by having the transiting dust clouds block a larger fraction of the circumstellar gas than of the white dwarf and by having all of them (transiting dust, circumstellar gas, and white dwarf) aligned with respect to our line of sight. 

The light curve of this object is extremely variable as shown by
Rappaport et al. \cite{2016MNRAS.458.3904R,2018MNRAS.474..933R} and G\"ansicke et al. \cite{gansicke16}.
This is because (i) individual objects have slightly different periods, (ii) the periods of some of individual objects can change slowly with time, and (iii) their dust activity can change dramatically on timescales of months and years.

High resolution spectroscopic observations also revealed the presence of high-velocity gas
orbiting the white dwarf. \cite{2016ApJ...816L..22X,2017ApJ...839...42R}.

A more detailed review of this object can be found in \cite{2018MNRAS.474..933R,2018haex.bookE..37V}. Very recently a second white dwarf with possibly related properties was discovered \cite{2019arXiv190809839V}.  This object, ZTF J013906.17+524536.89, exhibits two deep transits separated by 110 days, but it is not yet clear if this is a periodicity.

\subsection{Exo-comets}
\label{exocomet}

\begin{figure}
\centerline{
\includegraphics[angle=0,width=4.14cm]{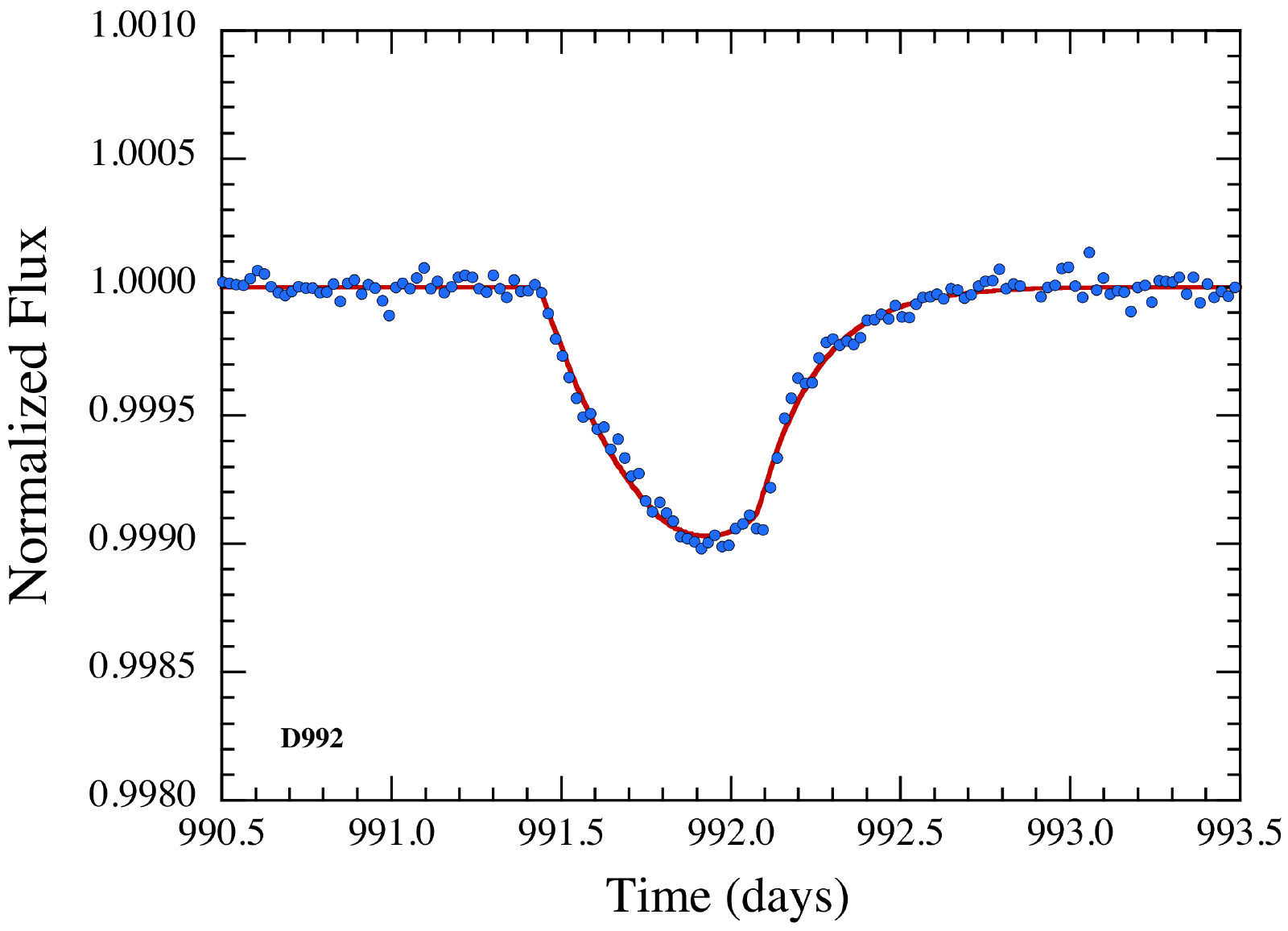}
\includegraphics[angle=0,width=3.6cm]{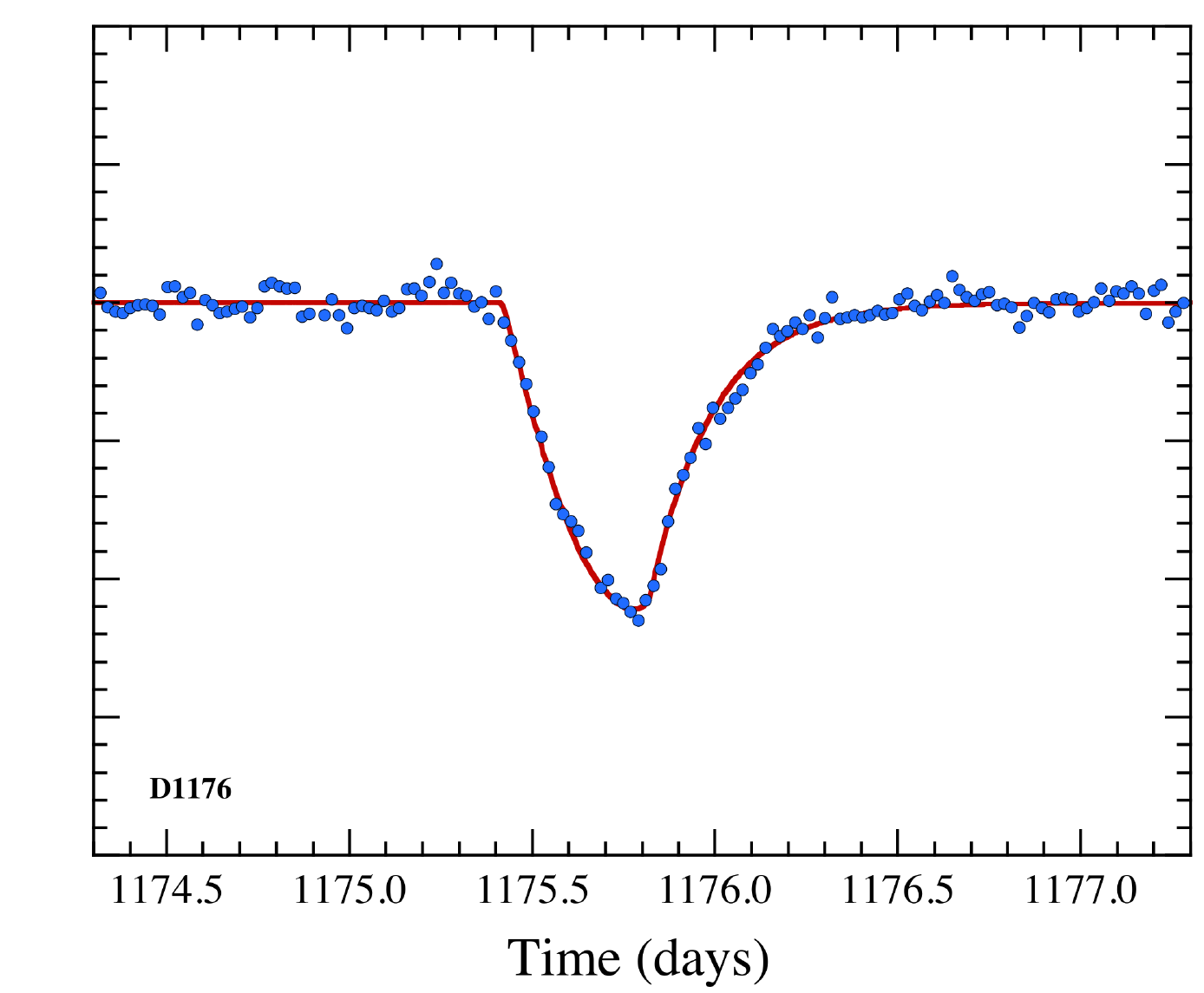}
\includegraphics[angle=0,width=3.6cm]{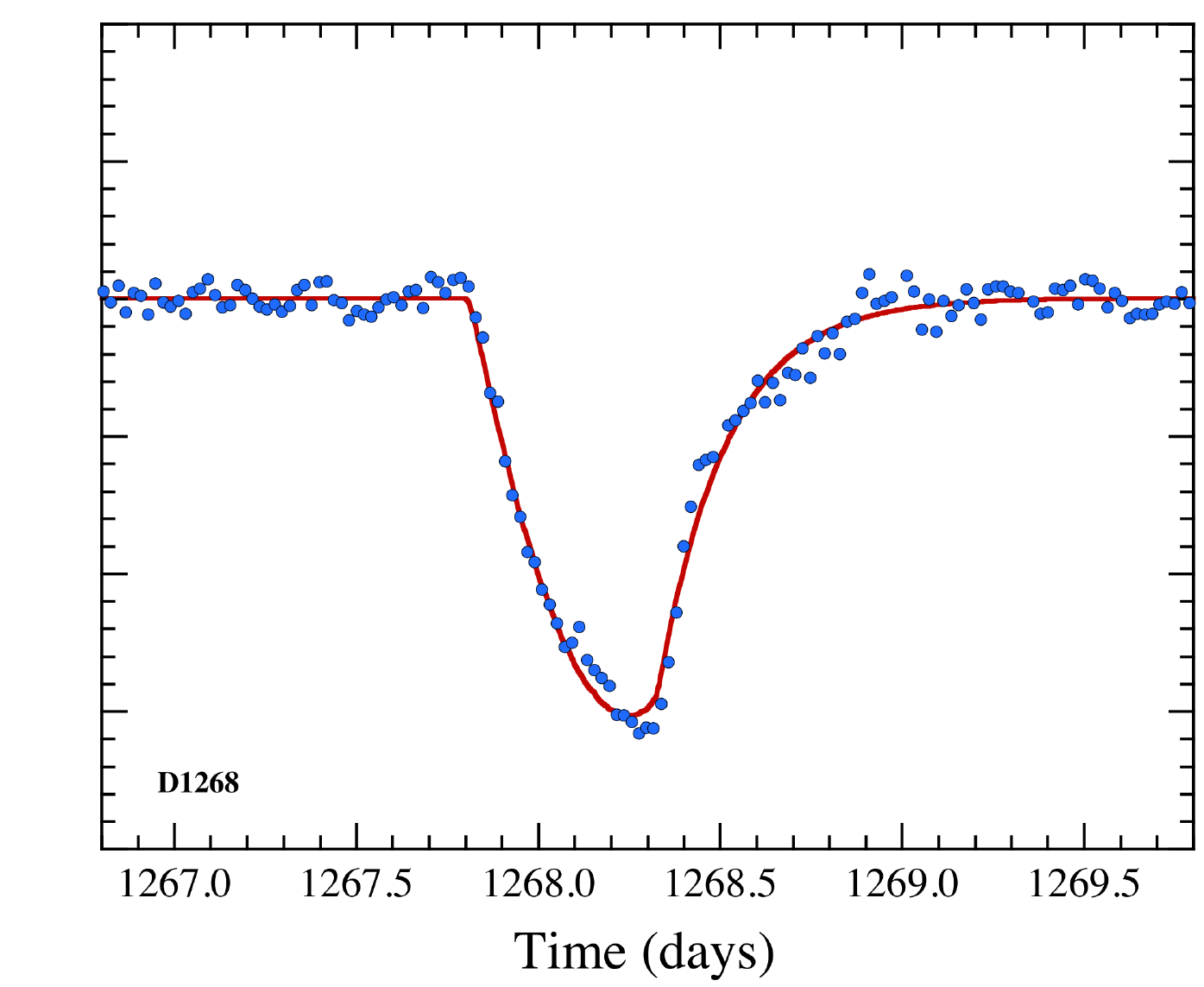}
}
\caption{
Three deeper transit events found in the Kepler light curve of KIC3542 by  \cite{rappaport18} by permission of Oxford University Press.}
\label{exocometfig}
\end{figure}

The unprecedented precision of the {\em Kepler} photometry enabled the detection of even smaller objects than planets or even large asteroids.
Two decades ago \cite{lecavelier99} predicted that comets orbiting other stars and emitting large dusty tails might be detected by photometry when transiting their host stars and calculated what their light curves could look like.  In the {\em Kepler} data,
\cite{rappaport18} detected six events in the light curve of KIC 3542116 (KIC3542) and one event in the light curve of KIC 11084727 (KIC1108) which looked very much like the expected cometary transits.
They were about 0.05-0.2\% deep and highly asymmetric, similar in shape
to the KIC1255b transits but several times more shallow (see Fig.\ref{exocometfig}). The three deeper transits of KIC3542 and that of KIC1108 lasted for about one day while the three shallower transits of KIC3542 lasted for about half a day.
There is no obvious periodicity to these events indicating that six transits
of KIC3542 are caused by 2-6 distinct comet-like bodies. 
The duration of the transits corresponds to a transverse speeds of about 35-50 km/s for the longer transits and about 75-90 km/s for the shorter transits.
This corresponds to orbital periods of $\geq 90$ and $\geq 50$ days, respectively.
Both host stars KIC3542 and KIC1108 are relatively bright (V=10mag) and hot stars
with $T_{\rm eff}=6900 ~{\rm and}~6800$K, respectively. Most of the stars monitored by {\em Kepler} are cooler (or older) so the fact that they are hotter/younger, similar to each other, and also similar to Boyajian's star mentioned later is probably not an accident. 

Recently, a single comet-like transit was found in the archival lightcurve of KIC 8027456 \cite{kennedy19}. The {\em TESS} mission also detected three other dips of this kind 
in $\beta$ Pictoris \cite{zieba19}.
Similar events were discovered in two stars (EPIC 205718330 and EPIC 235240266) monitored by the {\em K2} mission \cite{ansdell19}. 
The authors call these `little dippers' since they resemble the so-called ``dipper"
stars.
However, contrary to dipper stars. these dips are 1-2 orders of magnitude
shallower with depths of about 0.1-1\%. The dips in the `little dippers' are episodic, not periodic,
lasting for about 0.5-1 days, with complicated shapes resembling more WD1145 or Boyajian's star rather than the typical exocomet like profile seen in Fig.\ref{exocometfig}. Nevertheless, the authors argue that exocomets are the most likely explanation. The host stars are early-K and late-F type dwarfs, not
younger than 150 and 800 Myr, respectively.

Recently, two other 'dipper' stars were discovered by K2 mission.
HD139139 is a normal early G-type star which shows a sequence of 28 transit like dips \cite{rappaport19a}. 
The events are about 200 ppm deep, 0.7-7 hours long, are random, and do not show a significant asymmetry. 
EPIC204376071 is a young M5 dwarf which shows a single 80\% deep asymmetric dip \cite{rappaport19b}.
There is no explanation to these phenomena yet.


\section{Boyajian's star}
\label{tabby}

We would like to introduce another star which may be related
to the above mentioned objects and which is sometimes 
labeled as ``the most mysterious star in the Galaxy".

\subsection{Discovery and the Kepler light curve}

The {\em Kepler} mission delivered a huge amount of high-precision photometric light curves for about 170,000 stars.
A group of volunteers, the `Planet Hunters', were reviewing the light curves by human eye and they were the first to notice that there were some very strange dips in flux from  the star KIC 8462852.
A more detailed analysis and follow-up observations resulted in a discovery paper led by Tabetha Boyajian \cite{boyajian16}, and since that time the star has become known
as Boyajian's or Tabby's star.

\begin{figure}
\sidecaption[t]
\centerline{\includegraphics[width=11.cm]{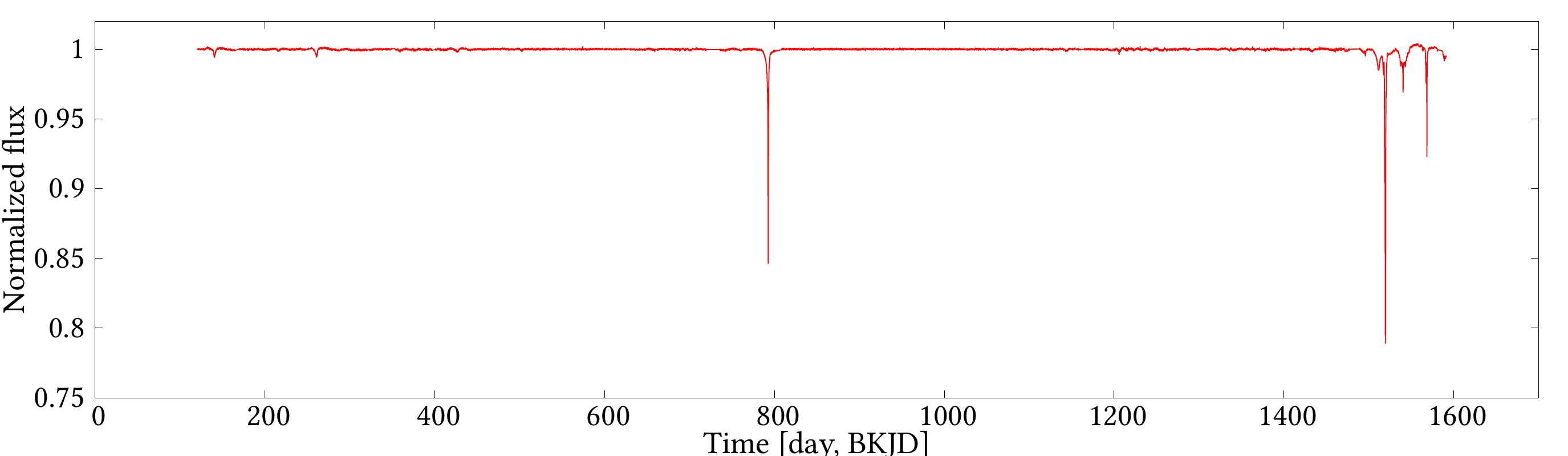}
}
\centerline{
\includegraphics[width=5.5cm]{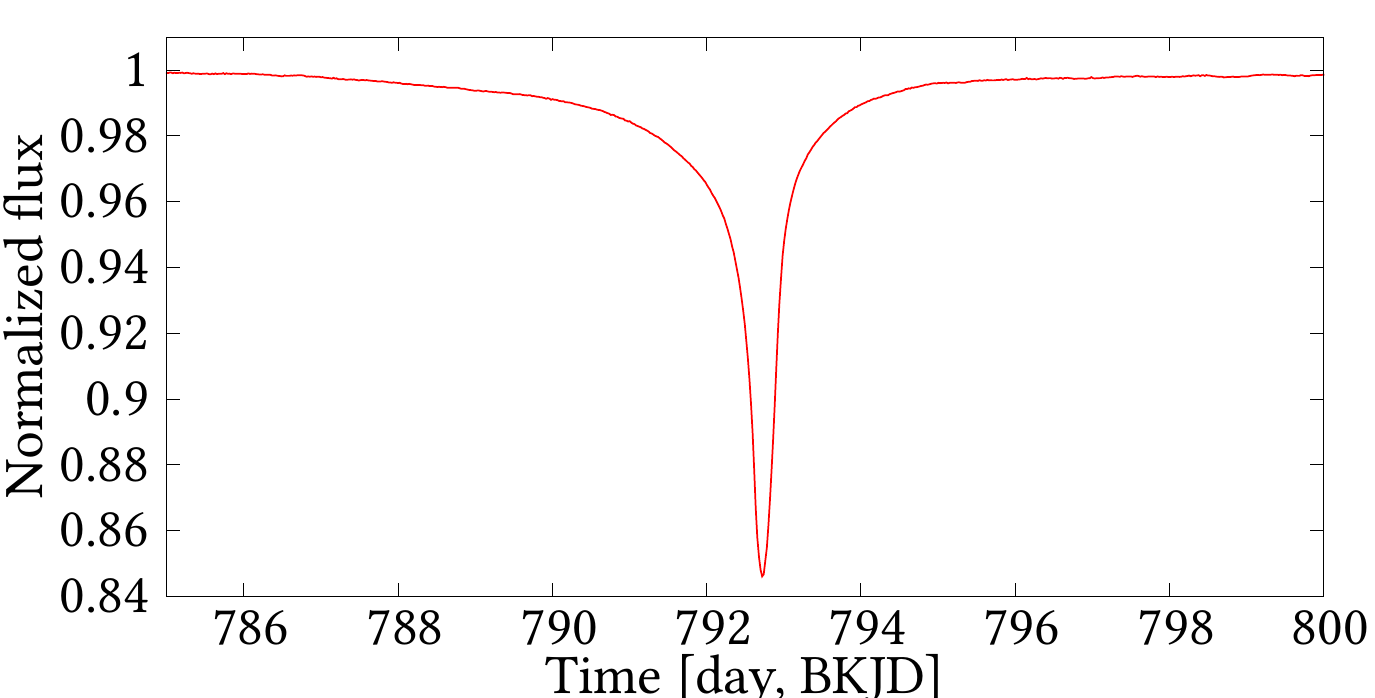}
\includegraphics[width=5.5cm]{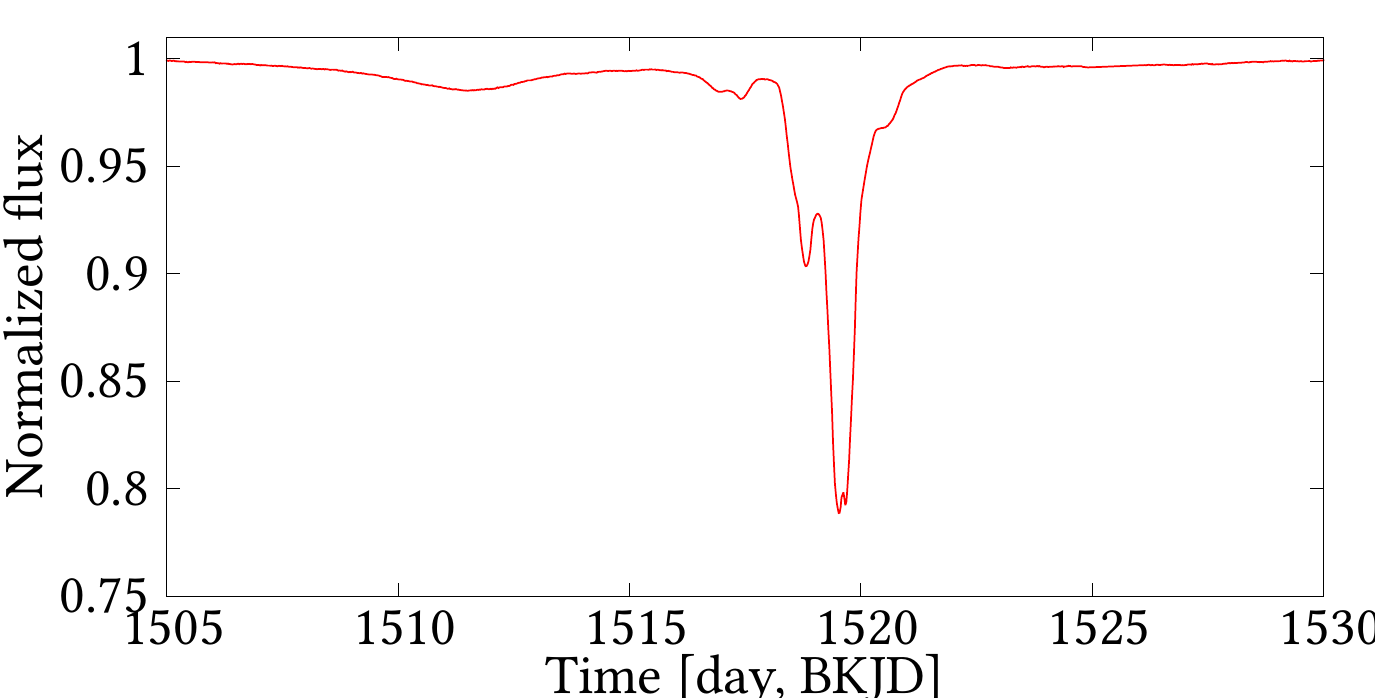}
}
\centerline{
\includegraphics[width=5.5cm]{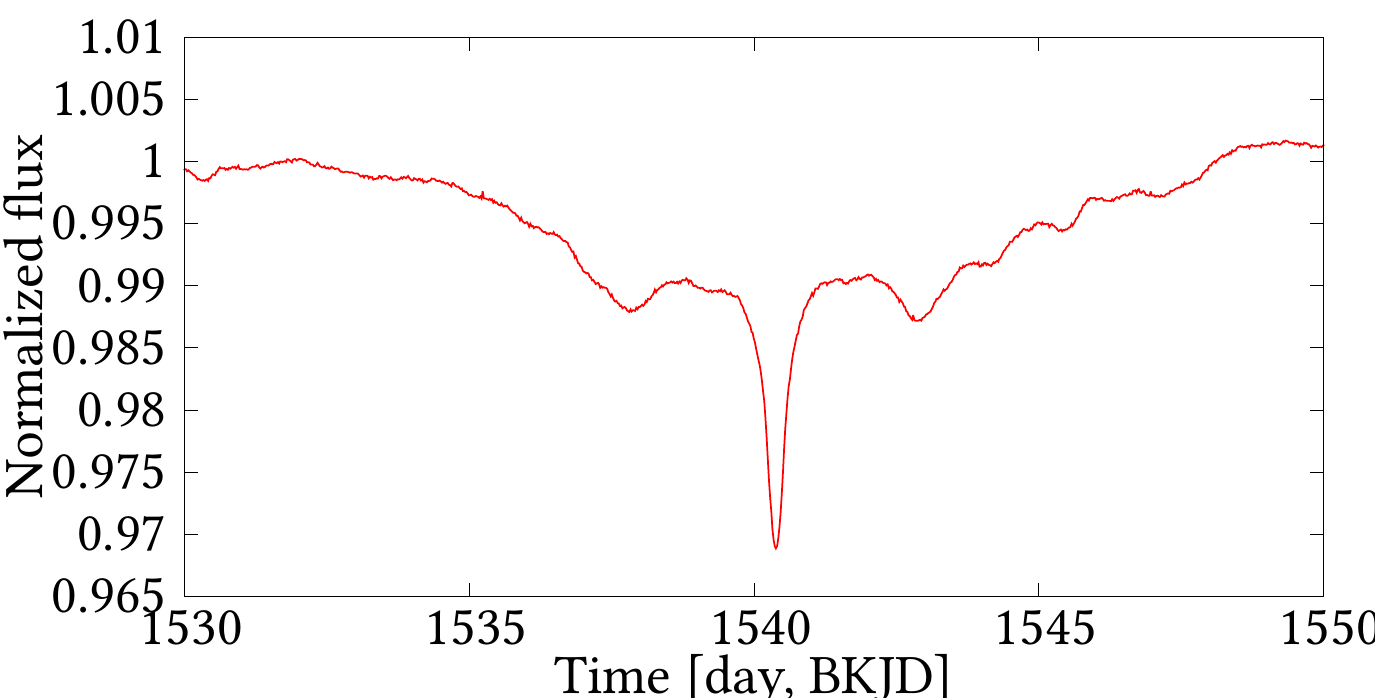}
\includegraphics[width=5.5cm]{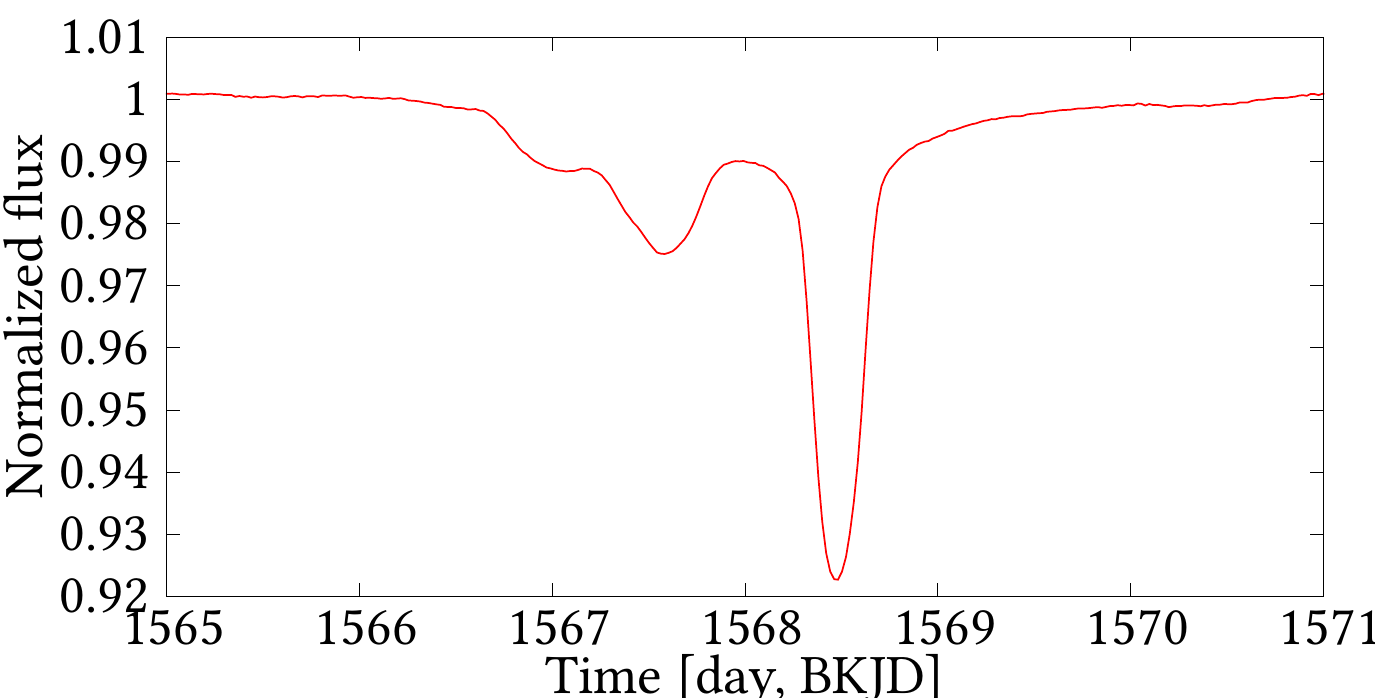}
}
\caption{
The {\em Kepler} light curve of Boyajian's star shows irregular dips (top).
A more detailed view of the four major events is displayed in the middle and bottom panels.
}
\label{tabbylc1}
\end{figure}

So what is so special about this star?
The {\em Kepler} light curve shows a few strong dimming events that are 10\%-20\% deep. They are irregular with no sign of periodicity and are clustered into four main events observed near BKJD=790, 1520, 1540, 1570 days\footnote{BKJD stands for the Kepler Barycentric Julian day which is a Julian Day minus 2454833}. They are shown in Figure \ref{tabbylc1}.
The D790 event is very smooth with a slow ingress followed by a faster egress.
The D1520 and D1570 events consist of a sequence of dips gradually increasing in strength.
D1540 is a symmetric triple dip with the central dip being the strongest.

There is another tiny feature in the Kepler data at D1210 which deserves attention.
It is a symmetric triple dip with the middle one being the strongest \cite{neslusan17}.
This shape resembles the D1540 event.

This kind of variability would not be anything unusual if this were a young star. Such stars are often surrounded by protostellar disks which might cause dipping events when seen nearly edge on. They show broad emission lines and infrared excess.
However, this star has no such features and looks like a normal F3V type main sequence star with temperature, mass, radius, projected equatorial velocity and rotational period of $T_{\rm eff}=6750$ K, $M=1.43 M_{\odot}$, $R=1.53 R_{\odot}$,  
$v \sin i=78$ km/s, and $P_{\rm rot}=0.88$ days, respectively \cite{boyajian16,martinez19}.\footnote{According to \cite{makarov16} the 0.88 day periodicity may come from a different source, not from the target star.}
It is a relatively bright, $V=11.7$ mag star at the distance of about 451 pc. The authors also discovered a faint M dwarf companion to the star at 1.95$''$ which is about 3.8 mag fainter in H band. However, this star is not physically bound to Boyajian's star \cite{clemens18}. 
%

\subsection{Follow up observations}

Soon after the discovery of Boyajian's star, a plethora of follow-up observations were performed.
Observations in the infrared region did not detect any infrared excess
but put constraints on the amount of dust at different distances from the star.
Spitzer/IRAC \cite{marengo15}, NASA/IRTF 3 m SpeX \cite{lisse15}, 
Millimetre (Submillimeter Array) and submillimetre (SCUBA-2) continuum observations
also did not detect any significant emission towards KIC 8462852. This places an upper limit of about $10^{-6} M_{\oplus}$ of dust lying within 2-8 AU from the star,
$10^{-3} M_{\oplus}$ located within 26 AU, and a total overall dust budget of <7.7 $M_{\oplus}$ within a radius of 200 AU \cite{thompson16}.

\begin{figure}
\sidecaption[t]
\includegraphics[width=7.cm]{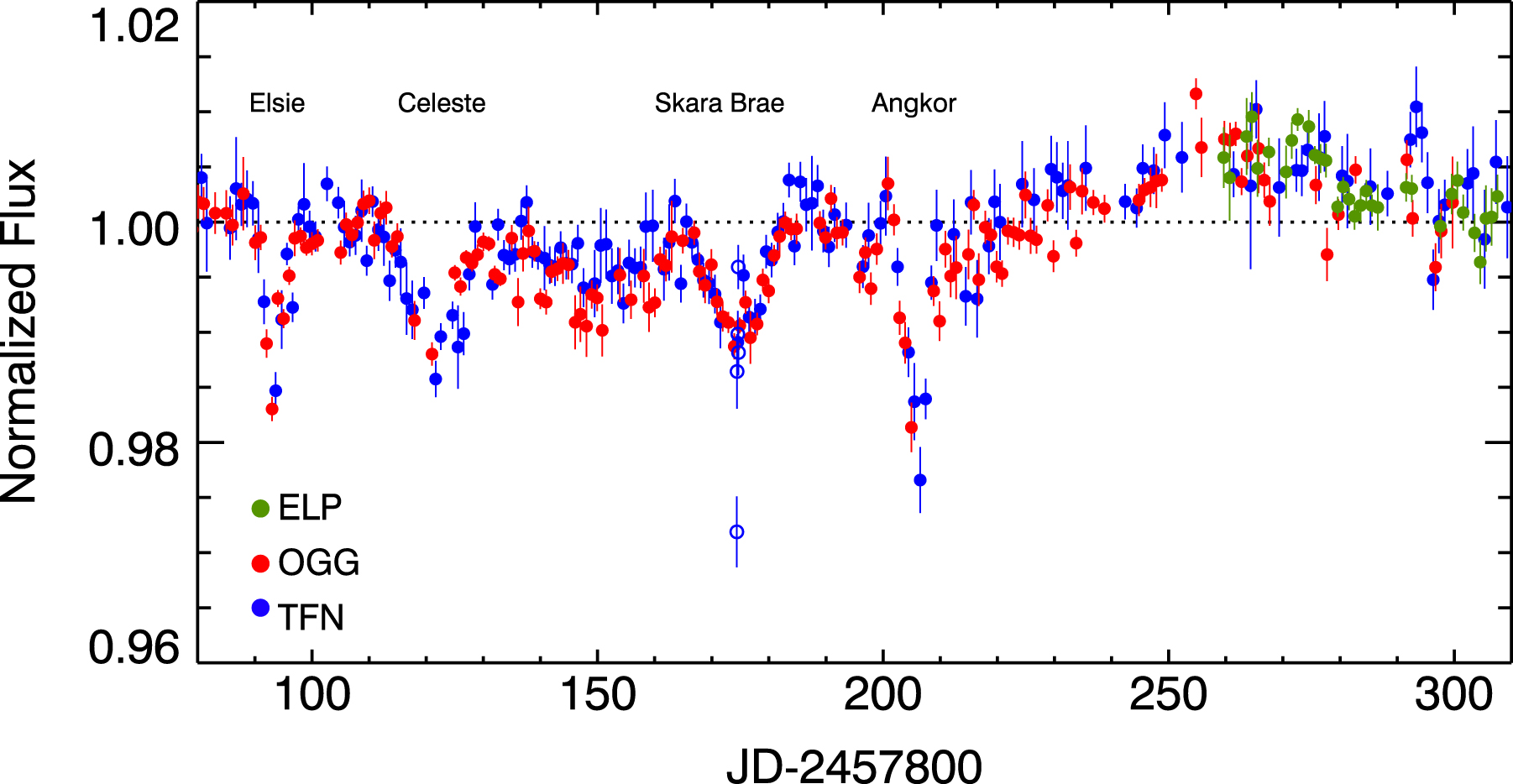}
\caption{
Boyajian's star became active again in May 2017. Ground based monitoring shows four
dips of depth 1-2.\%. Taken from \cite{boyajian18}, courtesy of ApJ.}
\label{groundlc1}
\end{figure}

Since the end of the {\em Kepler} space mission in 2013 May the star had been relatively quiet.
In 2017 May the dipping activity started again with four main events named `Elsie', `Celeste', `Skara Brae', and `Angkor' shown in Figure \ref{groundlc1} \cite{boyajian18}. These dips are about 1\%-2.5\% deep. The multiband photometry of the dips shows differential reddening favoring non-gray extinction. The data are inconsistent with dip models that invoke optically thick material, but rather they are in-line with predictions for an occulter consisting primarily of ordinary dust, where much of the material must be optically thin with particle sizes $\lesssim 1 \, \mu$m. 
No changes in the spectrum or polarization were detected during these events \cite{boyajian18,steele18,bodman19,martinez19}. 
Spectrophotometric observations of these recent dipping events with the GTC confirm that
the dips are deeper in the visual than at red wavelengths.
This is compatible with optically thin dust particles having sizes of $\simeq 0.0015-0.15 \,\mu$m. Such particles would be quickly blown away by the radiation pressure which indicates that the dust particles must be continuously replenished \cite{deeg18}. Finally, we note that the radial velocities of the host star also seem to be constant, within 2 sigma from the average value of $v_{\rm rad}=4.21\pm0.02$ km/s \cite{martinez19}.  This sets significant constraints on any companion stars or even brown dwarfs in short orbital periods.

\subsection{The long term variability}

There is evidence for a long term (secular) variability of Boyajian's star.
Based on archival photographic plates from Harvard College Observatory, \cite{schaefer16} found that
the star faded at an average rate of $0.164 \pm 0.013$ magnitudes per century from 1890 to 1989. This result was questioned by \cite{hippke16,hippke17}. Nevertheless, a similar study using archival photographic plates taken at the Maria Mitchell observatory during 1922-1991 found a similar trend of $0.12 \pm 0.02$ per century \cite{castelaz18}.

The star's brightness dropped significantly throughout the {\em Kepler} mission as well. Over the first 1000 days the star faded approximately by 0.9\%. 
It dimmed much more rapidly in the next 200 days, with its flux dropping by more than 2\% \cite{montet16}.
A slightly deeper 3.5\% drop was found in the contemporary GALEX observation in the near UV \cite{davenport18}. These results imply $R_{V} \simeq 5.0$ which, in turn, indicates circumstellar rather than interstellar dust attenuation.

Follow-up observations over a wide wavelength range from the UV to the mid-infrared from 2015 October through 2016 December, using Swift, Spitzer and AstroLAB IRIS indicate that the star faded in a manner similar to the long-term fading seen previously in {\em Kepler} data. According to \cite{meng17} the dimming rate for 
the entire period reported is "$22.1 \pm 9.7$ mmag per yr in the Swift wavebands, $21.0 \pm 4.5$ mmag in the ground-based B measurements, $14.0 \pm 4.5$ mmag in V, $13.0 \pm 4.5$ mmag in R, and $5.0 \pm 1.2$ mmag per yr averaged over the two warm Spitzer bands".
Continued ground based observations \cite{gary17,simon18} suggest that there are also brightening (not only fading) spells and that this long term variability may
be periodic with a period of 1600 days \cite{gary17}.
On the contrary, based on the ground and space based photometry spanning the 2006-2017 interval, it was concluded \cite{hippke18} that if the long term trend were to be periodic then the period would have to exceed 10 years.

\cite{wyatt18} developed a model of the dust cloud where the dust is distributed along a single elliptical orbit. They demonstrated that such a model satisfies the observational constraints set out by the lack of infrared excess and duration of the dips, and that it can explain the long term dimming. According to this model the dust must transit the star at 0.05-0.6 AU.
The ground based observations during 2015-2018 indicate that the long-term variability
is also chromatic. The amplitude is largest in the B band, while the VRI flux amplitudes are progressively smaller by factors of 0.77$\pm0.05$, 0.50$\pm0.05$, and 0.31$\pm0.05$, respectively \cite{schaefer18}. This implies that the dust particles causing the long-term variability must be about 0.1 $\mu$m in size. Such particles will be easily blown away and must be continuously replenished. The long-term variability (dimming) has a continuum of timescales ranging from almost a century, to decades, to years, and even down to a few months. It is most probably related to the shallow dip events and caused by the same phenomena.
The net result is that the star has experienced about a 12\% long-term dimming over the past century. This has serious implications for the amount of the dust that must be distributed along the elliptical orbit which now amounts to at least $10^{-3} M_{\oplus}$.

\subsection{Possible explanation and models}

There have been numerous models, ideas, and speculations proposed to explain the above mentioned behaviour. It is not possible to mention and discuss all the models here.
An overview was presented in \cite{wright16b} and it concluded that intervening interstellar material (ISM) is a more plausible explanation than other natural models. The discoverers themselves discussed a number of possibilities and favoured a comet scenario. Apart from that it was proposed that KIC 8462852 might be undergoing a late heavy bombardment, but is only in its very early stages \cite{lisse15,punzi18}.
It is also possible that the variability could be intrinsic to the star \cite{foukal17}, or
the dips might have been caused by matter in our Solar system \cite{katz17}.
According to \cite{metzger17} the secular dimming is the result of the inspiral of a planetary body or bodies into KIC 8462852, which took place $10-10^4$ yr ago.
The discoverers also proposed that the dips observed with {\em Kepler} may be due to transits of less massive bodies placed on eccentric orbits by the Lidov-Kozai oscillations due to the outer M-dwarf companion. However, the predicted smooth decline in flux is not in agreement
with the brightening episodes \cite{gary17,simon18}, and the M-dwarf companion turned
out not to be associated with Boyajian's star \cite{clemens18}.
However, evidence is growing that the dipping phenomenon is due to circumstellar dust.
In the next few sections we will mention three models that were developed
to the point where they can be directly compared with the observations of the dip events.

\subsubsection{A swarm of comets}
This is the scenario favoured by the discoverers and developed by \cite{bodman16}.
In this model the deep {\em Kepler} dips and long-term behaviour are due to transit of 
a large number (70-700) of comets. Such strings of comets are known from our Solar system so it is a natural explanation. An eccentric orbit has advantages. There is a high likelihood of the transits occurring near periastron and the material spends most of the time very far from the star so it can
satisfy the IR limits as well as the dynamical constraints \cite{wyatt18,schaefer18}. These comets must have had a common progenitor. 
The models can fit the {\em Kepler} dips very well.
Unfortunately, the model has a few drawbacks.
(i) It cannot reproduce the D790 event because it is very smooth and has a slow ingress and a faster recovery, while the model features just the opposite behaviour with a steeper ingress and a slower egress.
(ii) The symmetric triple dip, D1540, would require an accidental constellation of comets.
However, there are two other events of this kind: D1210 and Skara Brae, and they would have to be the result of an accidental grouping of comets as well.
(iii) Comets can hardly produce and continuously replenish $\gtrsim 10^{-3} M_{\oplus}$ of dust required to explain the long-term variability \cite{schaefer18}.
(iv) The model requires many free parameters (related to a large number of comets)
and even a perfect fit does not mean that it is correct. 

\subsubsection{Massive asteroids wrapped in dust}
This `recipe' can be found in \cite{neslusan17} but it was already considered in the
discovery paper \cite{boyajian16}.
It is in many respects similar to the above mentioned scenario.
According to this model there are a few massive asteroids or planetesimals surrounded by dust clouds
orbiting and transiting the star on eccentric orbits.
Obviously, the objects must have originated from a common progenitor as well.
The orbit and the amount of the dust required to transit the star is similar to the previous model so it also satisfies the IR limits and the dynamical constraints \cite{wyatt18,schaefer18}.
The difference is in the following.
Instead of a large number of comets only four more massive objects are sufficient to explain the four major Kepler events.
A massive object means that its gravity cannot be neglected, and it can retain a dust cloud within its Hill's sphere (contrary to a comet).
It naturally explains the smooth shape of the D790 event and produces a slower ingress and faster egress.
The symmetric triple dips: D1540, D1210, and Skara Brae are no longer due to an accidental
constellation of objects but rather single objects surrounded by dusty disks/rings.
The massive asteroids can produce and replenish $\gtrsim 10^{-3} M_{\oplus}$ of
dust to account for the long-term variability. It was demonstrated that if the objects were initially on exactly identical orbits, and were massive enough, then they (and their dust clouds) would mutually interact and end up on a slightly different orbits.
Even though the fits are not perfect, the model requires a small number of massive objects, and hence only a handful of free parameters.
One can anticipate that massive asteroids are accompanied by a large number of smaller debris which would account for the smaller dips and long-term variability.

\subsubsection{The Lord of the Rings}
We hope the reader will not mind the `label' above. The model was proposed by \cite{bourne18}.
The authors noticed that one of the post {\em Kepler} dips (Skara Brae, the one that occurred around Aug 9, 2017) is very similar to the {\em Kepler} D1540 event, i.e., it is a symmetric triple dip with the central dip being the deepest. The similarity is indeed striking and the authors presume that
it is the transit of the same body. This implies an orbital period of 1601 days.
`The Lord' is a dark and relatively massive object -- a brown dwarf orbiting
the star. It is accompanied by a `fellowship' of about 9 rings which
are about 0.2 AU across.
With this model the authors were able to reproduce the Skara Brae and D1540
events very well. Apart from that the model explains a tentative 1600 day
periodicity found in the long-term variability. The other dips observed by {\em Kepler}
were not modelled but might be understood assuming transits of additional bodies
(moons) related to the brown dwarf.
The model makes a very precise and testable prediction.
`The Return of the Lord' should happen during Christmas on Dec 27, 2021.

A similar idea was presented earlier in \cite{kiefer17}.
The authors identified two strikingly similar events in the {\em Kepler} light curve
which are approximately 0.1\% deep and occurred at D216 and D1144. They show that these events could be explained by the occultation of the star by a giant ring system or by the transit of a string of half a dozen exocomets. These events occurred 928.25 days apart and the authors predict that the next event will occur between 3-8 October 2019. 

More recent comparison and cross-correlation of {\em Kepler} dips and dips observed from the ground indicate a similar periodicity of 1574.4 days (4.31 yr) \cite{sacco18}.
This period also explains a few other historical dimming events of the star in the past. It predicts the next return of the D790 event on Oct 17, 2019.
We would like to comment that this idea presumes that the mutual gravitational
interaction among the bodies orbiting the star must be negligible. It is not compatible with the brown-dwarf hypothesis.

It remains to be established whether these models are compatible with the long-term variability, infrared limits, and various other constraints including the dynamics
of the system.


\section{Ongoing and future space missions}\label{future}

Because {\it Kepler} played a pioneering role in the detection of a new class of `disintegrating' objects, we shall briefly discuss ongoing and future space missions in order to present their potential for new discoveries of this particularly interesting class of objects.

\subsection{{\em TESS}}

\begin{figure}
\sidecaption[t]
\includegraphics[width=7.5cm,angle=0]{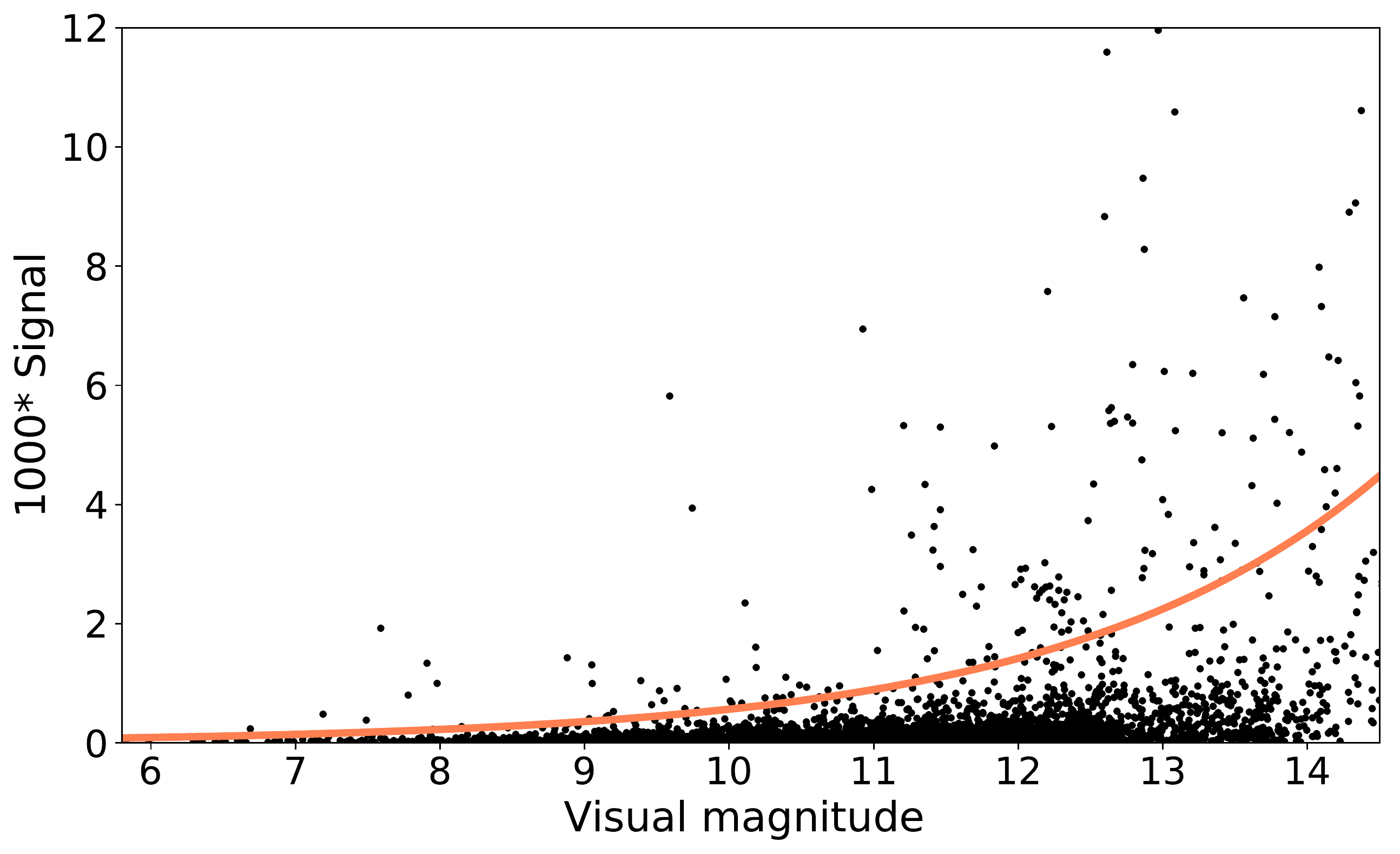}\caption{Expected distribution of {\em TESS} planets with the orange line representing a detection threshold for mid-sized telescopes. Figure from \cite{2019PASP..131h5001K}\,DOI: \href{https://iopscience.iop.org/article/10.1088/1538-3873/ab2143}{10.1088/1538-3873/ab2143}. \textcopyright The Astronomical Society of the Pacific 
Reproduced by permission of IOP Publishing.  All rights reserved.}\label{mst}
\end{figure}

{\em TESS} is a NASA space mission successfully launched in 2018 and planned for at least 2 years of operations \cite{ricker14,2015JATIS...1a4003R}. The aim of the {\em TESS} mission is the detection of several thousand exoplanets, mainly Neptune- and super Earth-sized. However, several hundred Jupiter-sized planet detections are expected as well. {\em TESS}
is delivering precise photometry down to about 200 ppm which is sufficient to detect a transiting super-Earth \footnote{https://heasarc.gsfc.nasa.gov/docs/tess/observing-technical.html}. The first {\em TESS planets} were recently announced \cite{2018A&A...619L..10G,2018ApJ...868L..39H}. 
There are many interesting objects discovered by {\em TESS}, such as a Neptune-sized planet HD 21749b with another, Earth-sized, planet HD 21749c in the same system 
\cite{2019ApJ...875L...7D} or the first {\em TESS} transiting brown dwarf \cite{Subjak}. Many of the {\em TESS} planets should be suitable for ground-based follow-up observations to detect exoplanetary atmospheres even with mid-sized telescopes \cite{2019PASP..131h5001K} as shown in Figure \ref{mst}. Furthermore, it is expected that {\em TESS} will detect additional interesting systems, and among those should be the types of disintegrating and dusty objects which we described in this review.  \newline

\subsection{The PLATO space mission}
 
The ESA M3 space mission PLATO (PLAnetary Transits and Oscillations of stars) will be launched in 2026. The PLATO space mission will consist of 26 telescopes monitoring large portion of sky (about 50\%) for transits with an unprecedented photometric accuracy of a few ppm \cite{rauer14}. The PLATO mission should find several thousand planetary candidates around one million bright stars from naked eye brightness to Vmag$=11$. PLATO will be able to detect even an Earth-like planet on an Earth-like orbit among the Solar type stars. PLATO will also focus on asteroseismology of stars \cite{2017EPJWC.16001003G}. However, the PLATO mission will also contribute to many other fields of astrophysics ranging from variable star research to extragalactic objects \cite{rauer14}. The majority of the PLATO targets and candidates will amenable to follow-up studies from the ground, thereby allowing for an exact determination of their masses and radii and thus allowing for their full characterization.

\subsection{The ARIEL space mission}

ARIEL, the Atmospheric Remote-sensing Infrared Exoplanet Large-survey, is an ESA M4 mission which will be launched in 2028 and it will be dedicated to unveiling the chemical composition of a sample of about 1000 selected transiting exoplanets \cite{tinetti12}. ARIEL will be equipped with an off-axis Cassegrain telescope with an elliptical primary mirror of 1.1-m $\times$ 0.7-m. ARIEL will be capable of photometric monitoring in visible and infrared wavelengths between 0.50-0.55 $\mu$m, 0.8-1.0 $\mu$m and 1.0-1.2 $\mu$m. A spectrograph with two medium resolving power channels  of 1.95-3.9\,$\mu$m and 3.9-7.8\,$\mu$m and one low-resolution channel of 1.25-1.95$\mu$m will be available \footnote{http://sci.esa.int/ariel/59798-summary/}. The precision of ARIEL should be sufficient to detect the signature of exo-atmospheres with a precision of at least $10^{-4}$ relative to the star. The main targets will be hot (600 K and more) planets, and it is expected that species like H$_2$O, CO$_2$, CH$_4$, NH$_3$, HCN or even metallic compounds such as TiO and VO will be detected and studied.

\begin{acknowledgement}
The authors would like to thank Prof. Saul Rappaport for carefully reviewing the manuscript and his help with several sections. We acknowledge the ERASMUS+ project 'Per aspera ad astra simul' under number 2017-1-CZ01-KA203-035562 which funded the mobilities of the authors. Apart from that JB thanks for support the VEGA 2/0031/18 and APVV 15-0458 grants. PK acknowledges the support of GACR grant number 17-01752J. E.P is partly financed by the Spanish Ministry of Economics and Competitiveness through projects ESP2016-80435-C2-1-R and PGC2018-098153-B-C31.
\end{acknowledgement}

\bibliographystyle{spphys}
\bibliography{references.bib,budaj2PK.bib,palle.bib}


\end{document}